\documentclass[aps,prx,twocolumn,showpacs]{revtex4-1}
\usepackage{bm}
\usepackage{mathrsfs}
\usepackage{amsmath}
\usepackage{amssymb}
\usepackage{graphicx}
\usepackage{amsfonts}
\usepackage{amsthm}
\usepackage{color}
\usepackage{dcolumn}
\usepackage{txfonts}
\usepackage[caption=false]{subfig}
\newcommand{\RN}[1]{%
  \textup{\uppercase\expandafter{\romannumeral#1}}%
}

\begin{document}

\title{Non-adiabatic transitions in parabolic and super-parabolic $\mathcal{PT}$-symmetric non-Hermitian systems}
\author{Chon-Fai Kam}
\author{Yang Chen}
\email{Corresponding author. Email: yangbrookchen@yahoo.co.uk}
\affiliation{Department of Mathematics, Faculty of Science and Technology,
University of Macau, Avenida da Universidade, Taipa, Macau, China}

\begin{abstract}
Exceptional points, the spectral degeneracy points in the complex parameter space, are fundamental to non-Hermitian quantum systems. The dynamics of non-Hermitian systems in the presence of exceptional points differ significantly from those of Hermitian ones. Here we investigate non-adiabatic transitions in non-Hermitian $\mathcal{P}\mathcal{T}$-symmetric systems, in which the exceptional points are driven through at finite speed which are quadratic or cubic functions of time. We identity different transmission dynamics separated by exceptional points, and derive analytical approximate formulas for the non-adiabatic transmission probabilities. We discuss possible experimental realizations with a $\mathcal{P}\mathcal{T}$-symmetric non-Hermitian one-dimensional tight-binding optical waveguide lattice.  
\end{abstract}

\maketitle

\section{introduction}\label{I}
In recent years, the emerging field of non-Hermitian quantum systems with parity-time ($\mathcal{P}\mathcal{T}$) symmetry \cite{bender1998real, bender1999pt, heiss1990avoided, heiss2000repulsion, bender2005introduction, bender2007making, moiseyev2011non, bender2018pt, kato2013perturbation, longhi2017oscillating, gong2018piecewise, zhang2018dynamically, zhang2018hybrid, longstaff2019nonadiabatic}, i.e., non-isolated coupled quantum systems with balanced gain and loss \cite{feng2017non, longhi2018parity, el2018non, ozdemir2019parity}, has attached great interest due to the potential for new quantum devices and applications \cite{el2007theory, makris2008beam, klaiman2008visualization, guo2009observation, ruter2010observation, brandstetter2014reversing, peng2014loss, chang2014parity, lin2011unidirectional, regensburger2012parity, feng2013experimental, hodaei2014parity, feng2014single, longhi2010pt, chong2011p, sun2014experimental, wiersig2014chiral, kim2014partially, peng2016chiral, miao2016orbital}, and has opened up new opportunities and challenges for both theorists and experimentalists \cite{gao2015observation, peng2016anti, zhang2016observation, xu2016topological, jing2014pt, jing2015optomechanically, jing2017high, zhang2018phonon, schonleber2016optomechanical, zhu2014p, fleury2015invisible, ding2016emergence, schindler2011experimental, benisty2011implementation, alaeian2014parity, kang2013effective, kang2016chiral, xiao2016effective, fleury2014negative, ding2015coalescence, zhen2015spawning, cerjan2016exceptional, konotop2016nonlinear, suchkov2016nonlinear}.

In conventional quantum mechanics, it is an axiom that the dynamics of a state of an isolated quantum system is governed by a Hermitian Hamiltonian ($\hat{H}=\hat{H}^\dagger$), which ensures real energy eigenvalues as well as an unitary time evolution for which the total probability of finding a particle in space is conserved. However, for a non-isolated quantum system with gain and loss, the total probability is in general not conserved, which yields a non-unitary time evolution described by a non-Hermitian Hamiltonian ($\hat{H}\neq \hat{H}^\dagger$). Remarkably, as shown by Bender and Boettcher \cite{bender1998real, bender1999pt}, the realness of the eigenvalues is ensured by a wide class of non-Hermitian Hamiltonians, i.e., class of Hamiltonians which are symmetric under the parity-time ($\mathcal{P}\mathcal{T}$) transformations. Here, the actions of the parity $\mathcal{P}$ and time $\mathcal{T}$ operators are defined as $\mathcal{P}$: $i\rightarrow i$, $\hat{x}\rightarrow-\hat{x}$, $\hat{p}\rightarrow-\hat{p}$ and $\mathcal{T}$: $i\rightarrow -i$, $\hat{x}\rightarrow\hat{x}$, $\hat{p}\rightarrow-\hat{p}$, where $\hat{x}$ and $\hat{p}$ are position and momentum operators respectively \cite{bender1998real, bender1999pt}. Hence, the action of the parity-time operator is $\mathcal{P}\mathcal{T}$: $i\rightarrow-i$, $\hat{x}\rightarrow -\hat{x}$, $\hat{p}\rightarrow\hat{p}$, where the operator $\mathcal{P}$ is linear, and the operator $\mathcal{T}$ is anti-linear, as it changes the sign of $i$. The operators $\mathcal{P}$ and $\mathcal{T}$ are commute, i.e., $[\mathcal{P},\mathcal{T}]=0$, which satisfy the relations $\mathcal{P}^2=\mathcal{T}^2=1$, $\mathcal{P}=\mathcal{P}^\dagger$ and $\mathcal{T}=\mathcal{T}^\dagger$ \cite{bender2005introduction}. For a single particle in one-dimensional space equipped with a Hamiltonian $\hat{H}\equiv \hat{p}^2/2m+\hat{V}(x)$, the condition of $\mathcal{P}\mathcal{T}$ symmetry is equivalent to $\hat{V}(x)=\hat{V}^*(-x)$.

Similar to the connection between symmetries and degeneracies of energy levels in Hermitian systems, $\mathcal{P}\mathcal{T}$ symmetries lead to a new type of spectral degeneracies in non-Hermitian systems, known as the exceptional points, where the real and imaginary parts of certain eigenvalues, as well as the associated eigenstates coalesce. In a striking contrast to the spectral degeneracies of Hermitian Hamiltonian, at which the eigenstates can still be chosen to be orthogonal to one another, the spectral degeneracies induced by $\mathcal{P}\mathcal{T}$ symmetries cause a loss of dimensions in non-Hermitian systems, as certain eigenstates become completely parallel and the Hamiltonian matrix becomes defective at the exceptional points. In other words, exceptional points are branch point singularities of the spectrum of the Hamiltonian \cite{kato2013perturbation}, at which certain eigenstates posses a finite overlap even in the absence of any perturbation. 

The intriguing properties of $\mathcal{P}\mathcal{T}$ symmetric non-Hermitian systems give rise to many counterintuitive features. A general $\mathcal{P}\mathcal{T}$ symmetric Hamiltonian may undergo a parity-time symmetry breaking phase transition, in which complex eigenvalues appear. For a non-Hermitian two-state Hamiltonian $\hat{H}$ with eigenstates $\phi_1$ and $\phi_2$, and eigenvalues $\lambda_1\neq\lambda_2$, the condition of $\mathcal{P}\mathcal{T}$ symmetry leads to $\hat{H}(\mathcal{P}\mathcal{T}\phi_1)=\lambda_1^*(\mathcal{P}\mathcal{T}\phi_1)$, and similarity for $\phi_2$. Clearly, the states $\mathcal{P}\mathcal{T}\phi_{1,2}$ are also eigenstates of $\hat{H}$ with eigenvalues $\lambda^*_{1,2}$. Hence, the simplest solution is $\mathcal{P}\mathcal{T}\phi_{1,2}=\phi_{1,2}$ and $\lambda_{1,2}^*=\lambda_{1,2}$, indicating the realness of the eigenvalues. However, there always exists another possible solution, $\mathcal{P}\mathcal{T}\phi_{1,2}=\phi_{2,1}$ and $\lambda_{1,2}^*=\lambda_{2,1}$, which shows that $\phi_1$ and $\phi_2$ are no longer the simultaneous eigenstates of the $\mathcal{P}\mathcal{T}$ operator, and the associated eigenvalues form a complex conjugate pair. In this regard, even though the Hamiltonian still possesses the $\mathcal{P}\mathcal{T}$ symmetry, it is spontaneously broken in certain regions of the parameter space, accompanied with complex eigenvalue bifurcation. By changing the parameters, one may reveal the underlying eigenvalue topological structure of non-Hermitian systems, where the real and imaginary parts of the eigenvalues form a set of multi-sheet Riemann surfaces centered around the exceptional points in the parameter space. When encircling an exceptional point, there is an unconventional level crossing behavior, accompanied with a phase change of one eigenstate but not of the other \cite{heiss1990avoided, heiss2000repulsion}. Particularly intriguing behavior is that dynamically encircling an exceptional point leads to chiral behaviors \cite{longhi2020non}, such that the encircling direction of the exceptional point determines the final output state \cite{doppler2016dynamically}.

In this work, we consider the dynamics of a non-Hermitian $\mathcal{P}\mathcal{T}$ symmetric system which directly goes through an assembly of exceptional points. Despite of its great importance, the non-Hermitian generalization of the two-level Landau-Zener paradigm has only recently been analyzed by Longstaff \cite{longstaff2019nonadiabatic}, and the associated non-Hermitian Landau-Zener-St\"{u}ckelberg interferometry was analyzed by Shen \cite{shen2019landau}. Here, we go one step further and analyze, both analytically and numerically, the non-hermitian generalization of the parabolic and super-parabolic models, in which the exceptional points are driven through at finite speed which are quadratic or cubic functions of time. We consider the case that the system is almost Hermitian when the parameters are far away from the exceptional points, such that the instantaneous eigenstates are nearly orthogonal. Specifically in this case it is relevant to consider the transmission probabilities that are the ratio of the transmission populations to the total population. We derive analytical approximate formulas for the transmission populations as well as the transmission probabilities. Unlike previous studies on non-adiabatic transitions in Hermitian systems \cite{kam2020analytical, kam2019analytical}, in which the transition points separating different dynamics regions are not predetermined, the benefit of our approach is to approximate separately the transmission dynamics by simple functions like hyperbolic or hypergeometric ones in the predetermined regions of broken and unbroken $\mathcal{P}\mathcal{T}$ symmetry. 

\section{Non-adiabatic transitions in Non-Hermitian two-level systems}\label{II}
To begin with, let us consider the following simple 2 $\times$ 2 non-Hermitian Hamiltonian matrix
\begin{equation}\label{NonHermiteHamiltonian}
    \hat{H} = \left( {\begin{array}{cc}
   -v & i\Gamma \\
   i\Gamma & v \\
  \end{array} } \right),
\end{equation}
\begin{figure}[tbp]
\begin{center}
\subfloat[$v(t)=t+t^2,\Gamma=1$\label{sfig:Fig_7a}]{%
  \includegraphics[width=0.48\columnwidth]{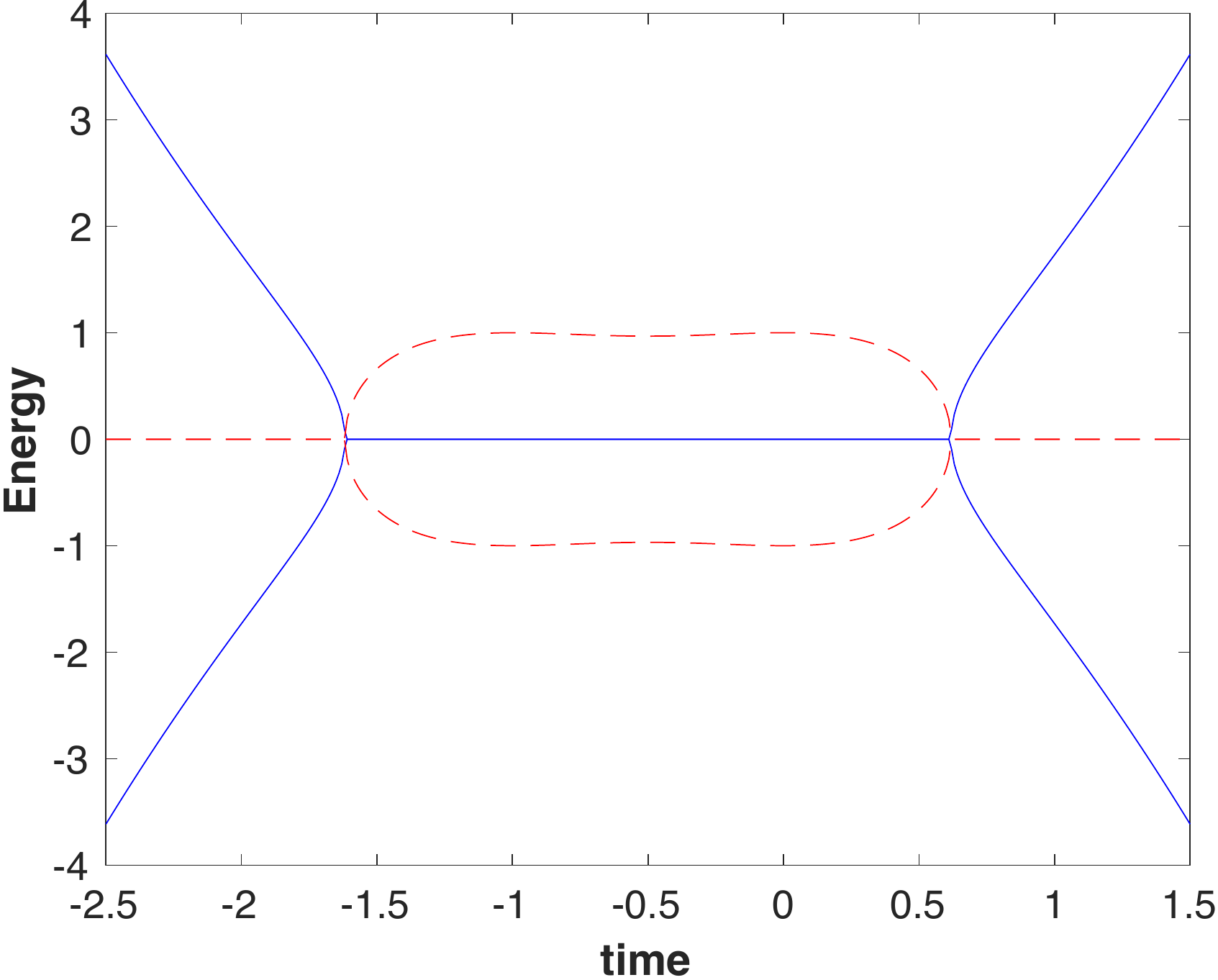}%
}\hfill
\subfloat[$v(t)=t+t^2,\Gamma=1$\label{sfig:Fig_7b}]{%
  \includegraphics[width=0.48\columnwidth]{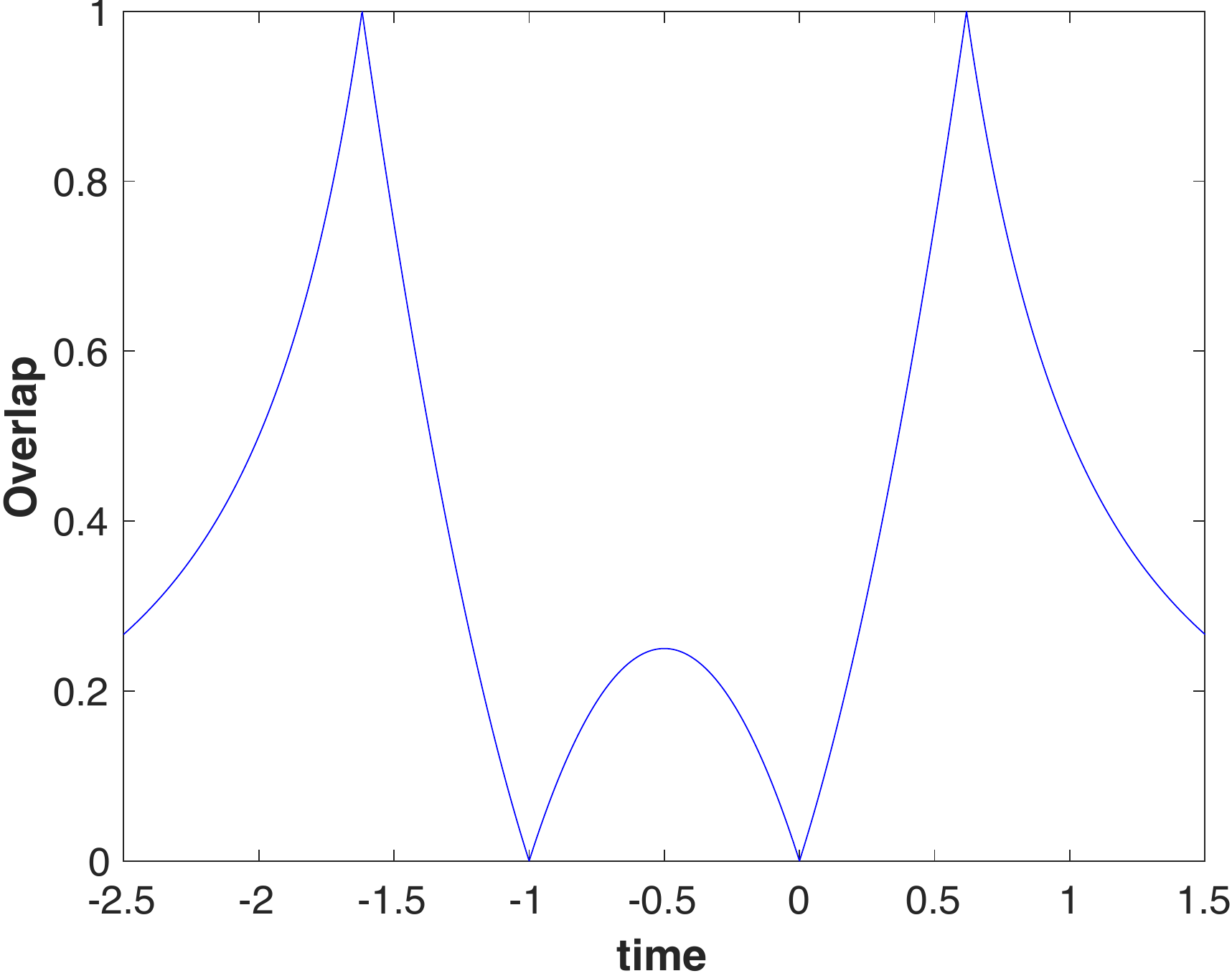}%
}\hfill
\subfloat[$v(t)=\frac{5t}{4}+\frac{t^2}{4},\Gamma=1$\label{sfig:Fig_7c}]{%
  \includegraphics[width=0.48\columnwidth]{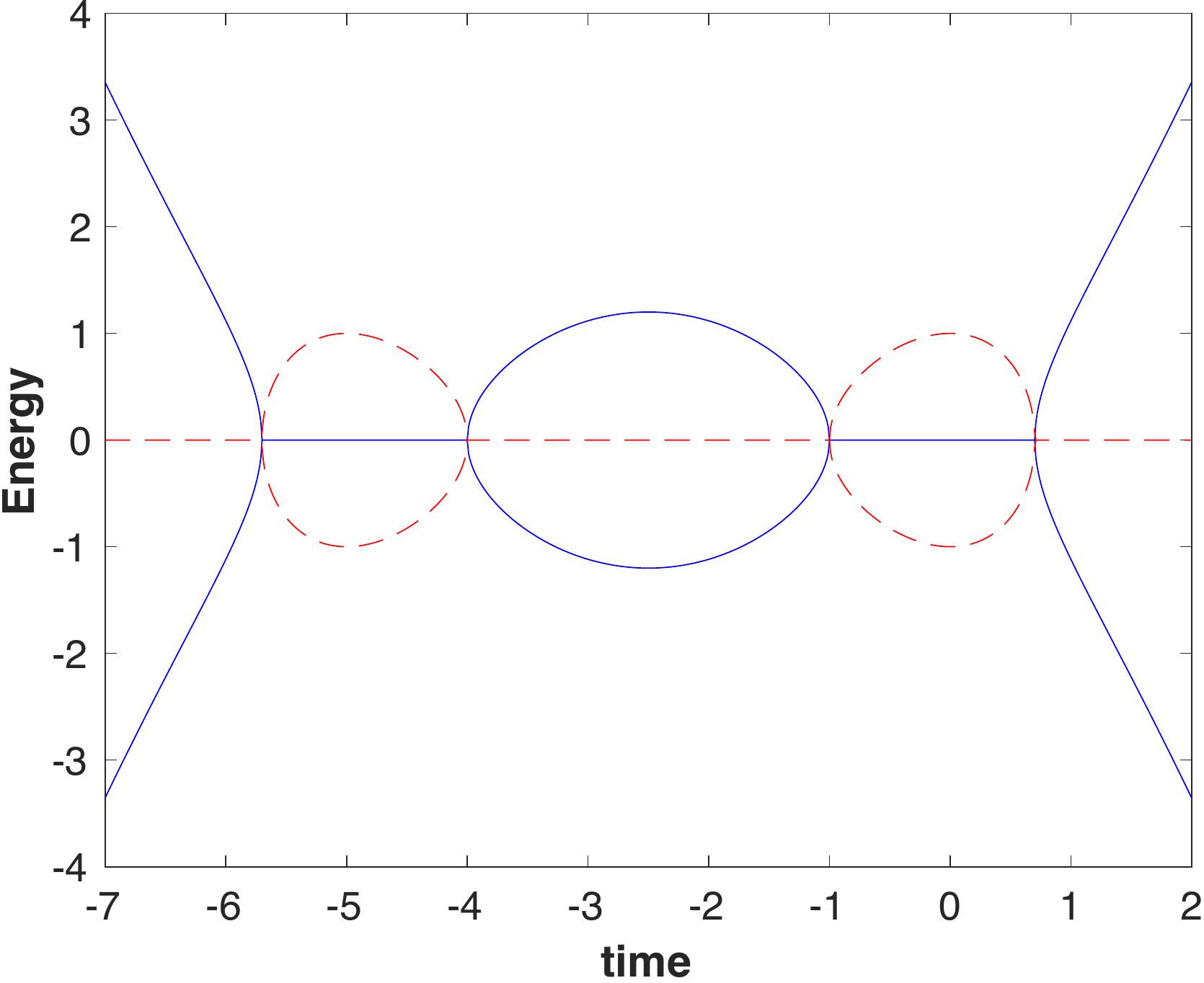}%
}\hfill
\subfloat[$v(t)=\frac{5t}{4}+\frac{t^2}{4},\Gamma=1$\label{sfig:Fig_7d}]{%
  \includegraphics[width=0.48\columnwidth]{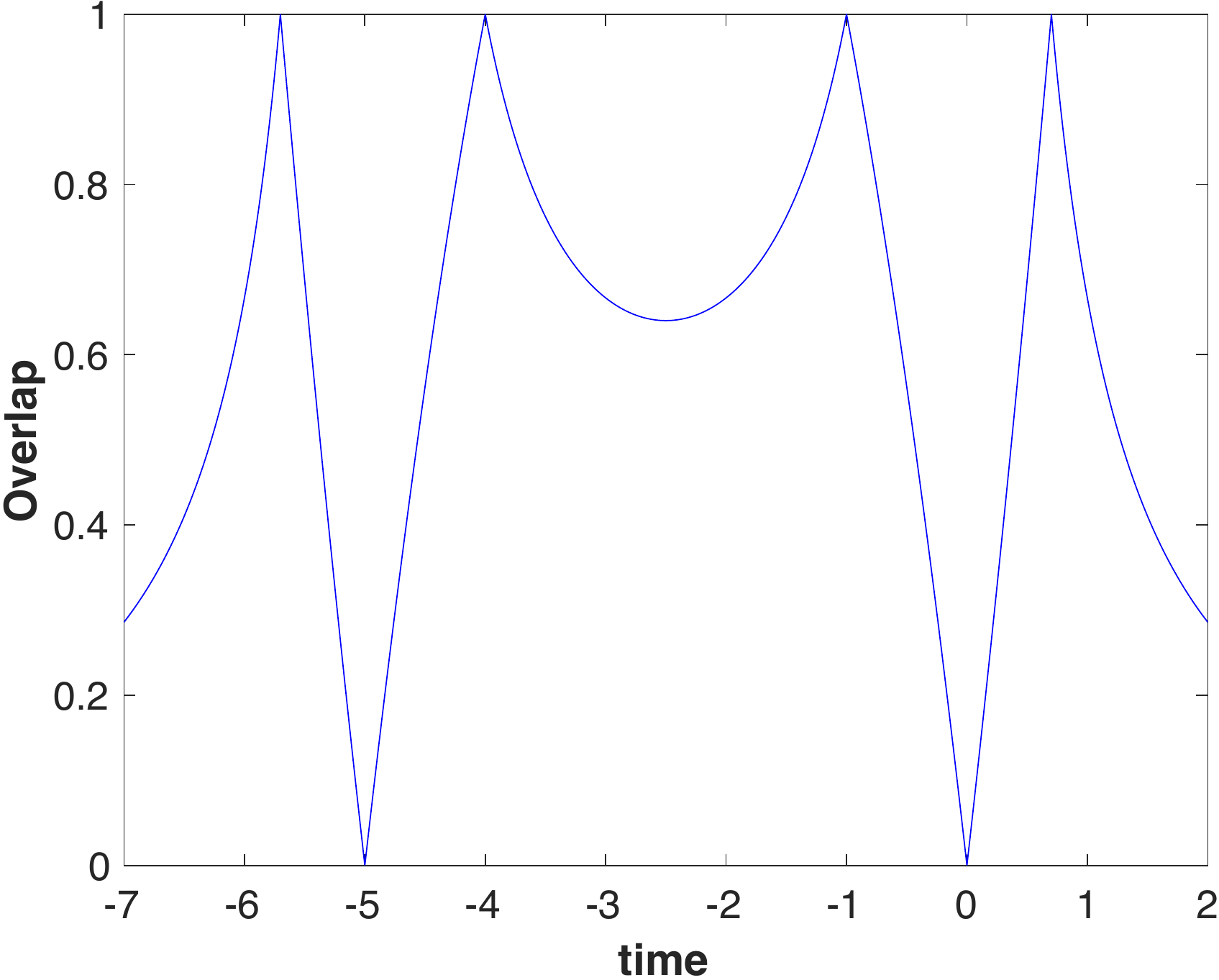}%
}
\caption{Schematic of the energies and overlap of the adiabatic eigenstates for the parabolic case with $v(t)\equiv\alpha t+\beta t^2$ and $\Gamma=\mbox{const}$. The real and imaginary parts of the eigen-energies are shown in blue solid lines and red dash lines respectively.}
\label{fig:NonHermitianModel2}
\end{center}
\end{figure}which is $\mathcal{P}\mathcal{T}$-symmetric, i.e., $[\mathcal{P}\mathcal{T},\hat{H}]=0$, where $\mathcal{P} \equiv \bigl( \begin{smallmatrix}0 & 1\\ 1 & 0\end{smallmatrix}\bigr)$ and $\mathcal{T}$ performs complex conjugation \cite{bender2007making}. Here, the parameters $v$ and $\Gamma$ are all real, and $\Gamma$ can be set to positive without loss of generality. This Hamiltonian describes two states with an energy difference $2v$ and a nonreciprocal coupling $i\Gamma$. Because of the non-hermiticity, there are two different regions in the parameter space. For $|v|>\Gamma$, the two eigenvalues $\epsilon_\pm\equiv\pm\sqrt{v^2-\Gamma^2}$ are all real, which refers to the region of unbroken $\mathcal{P}\mathcal{T}$ symmetry. On the other hand, for $|v|<\Gamma$, the two eigenvalues $\epsilon_\pm\equiv\pm i\sqrt{\Gamma^2-v^2}$ are complex conjugate to each other, which refers to the region of broken $\mathcal{P}\mathcal{T}$ symmetry. The boundaries of the two regions, i.e, $|v|=\Gamma$, are an assembly of exceptional points at which the two eigenvalues coalesce, and the associated eigenstates become completely parallel. 

We may calculate the left and right instantaneous adiabatic eigenstates $|\chi_{\pm}\rangle$ and $|\phi_{\pm}\rangle$, which are defined by $\hat{H}|\phi_{\pm}\rangle=\epsilon_{\pm}|\phi_{\pm}\rangle$ and $\langle\chi_{\pm}|\hat{H}=\epsilon_{\pm}\langle\chi_{\pm}|$, and can be explicitly expressed as
\begin{figure}[tbp]
\begin{center}
\subfloat[$v(t)=t+t^3,\Gamma=1$\label{sfig:Fig_1a}]{%
  \includegraphics[width=0.48\columnwidth]{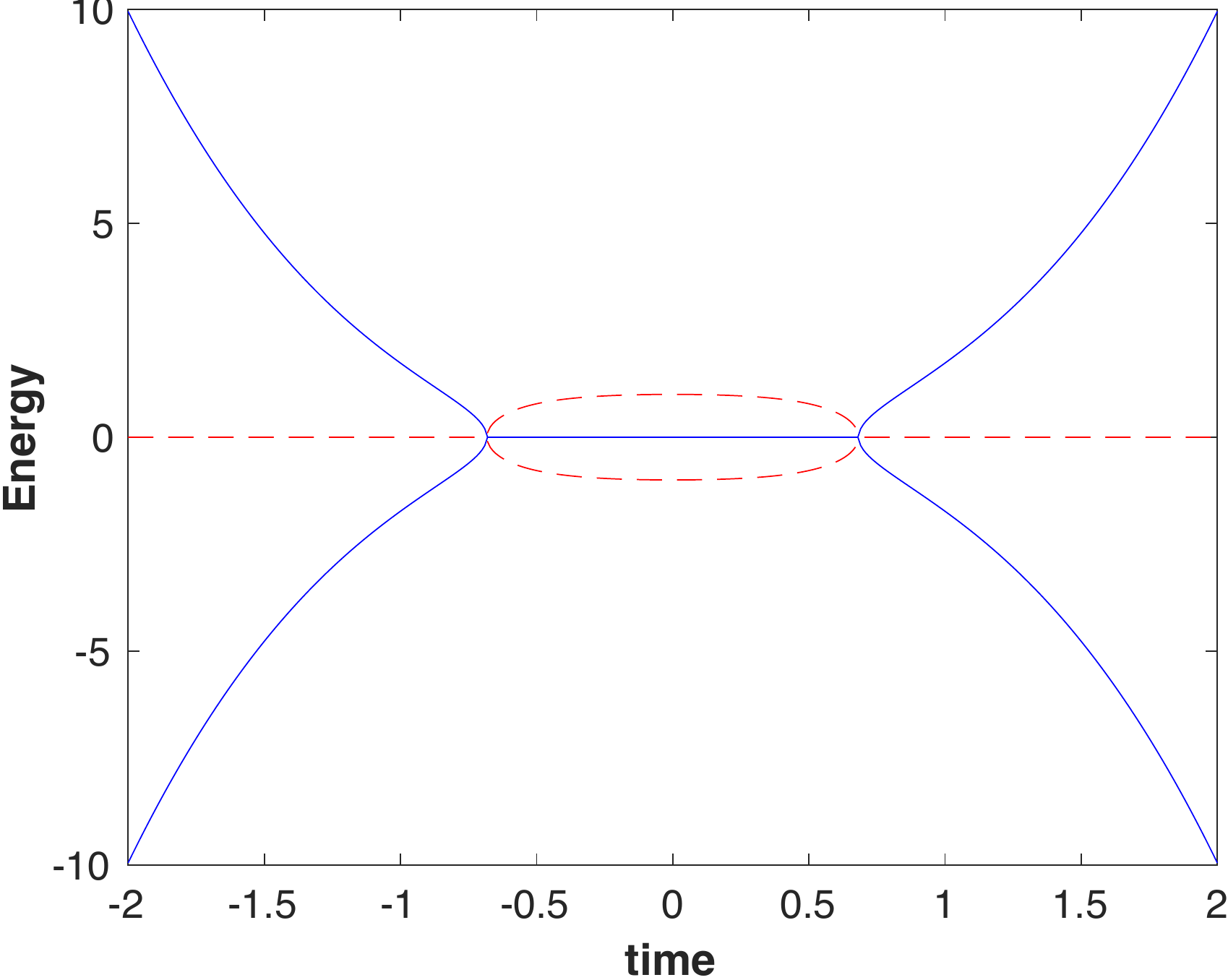}%
}\hfill
\subfloat[$v(t)=t+t^3,\Gamma=1$\label{sfig:Fig_1b}]{%
  \includegraphics[width=0.48\columnwidth]{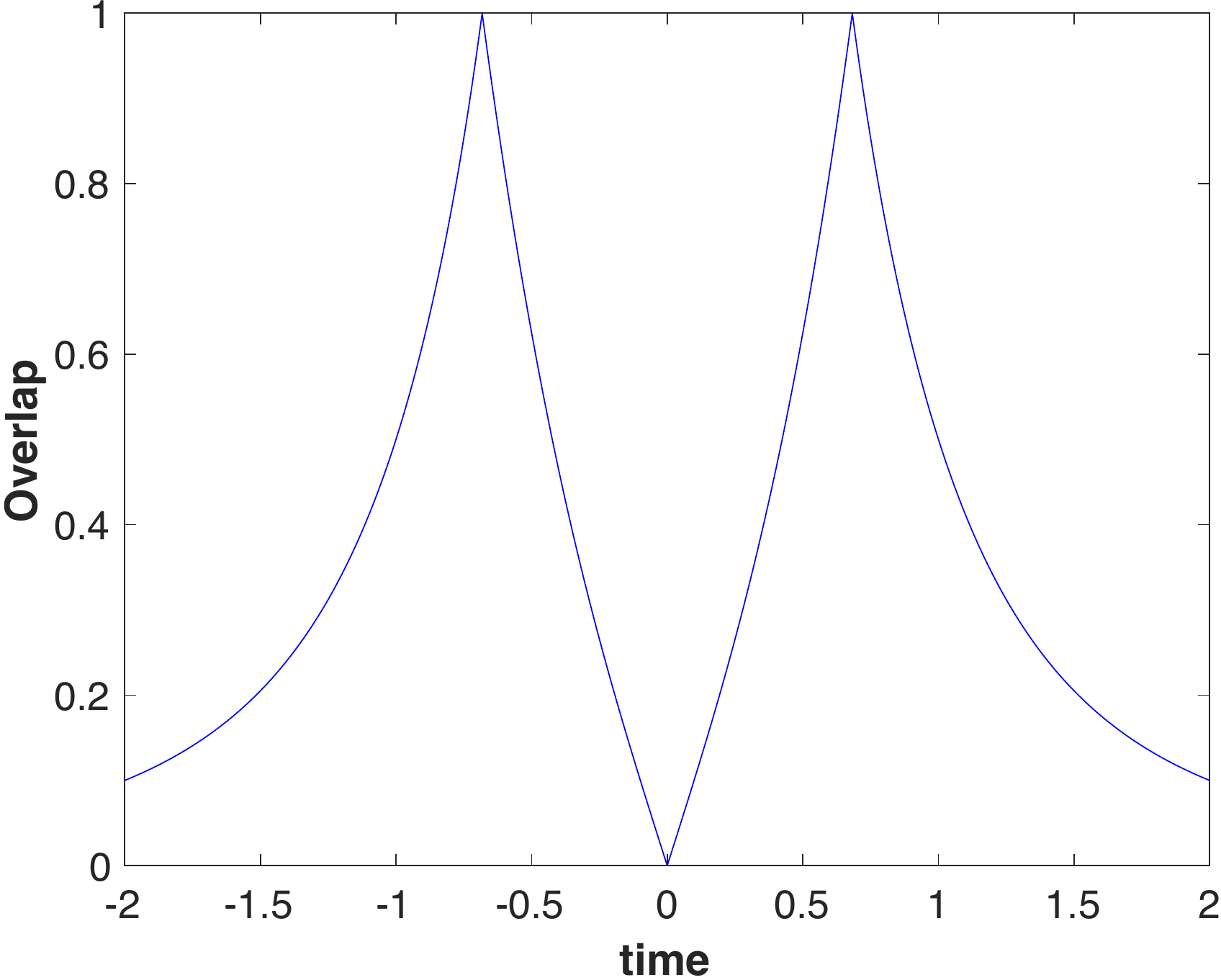}%
}\hfill
\subfloat[$v(t)=-2t+0.35t^3,\Gamma=1$\label{sfig:Fig_1c}]{%
  \includegraphics[width=0.48\columnwidth]{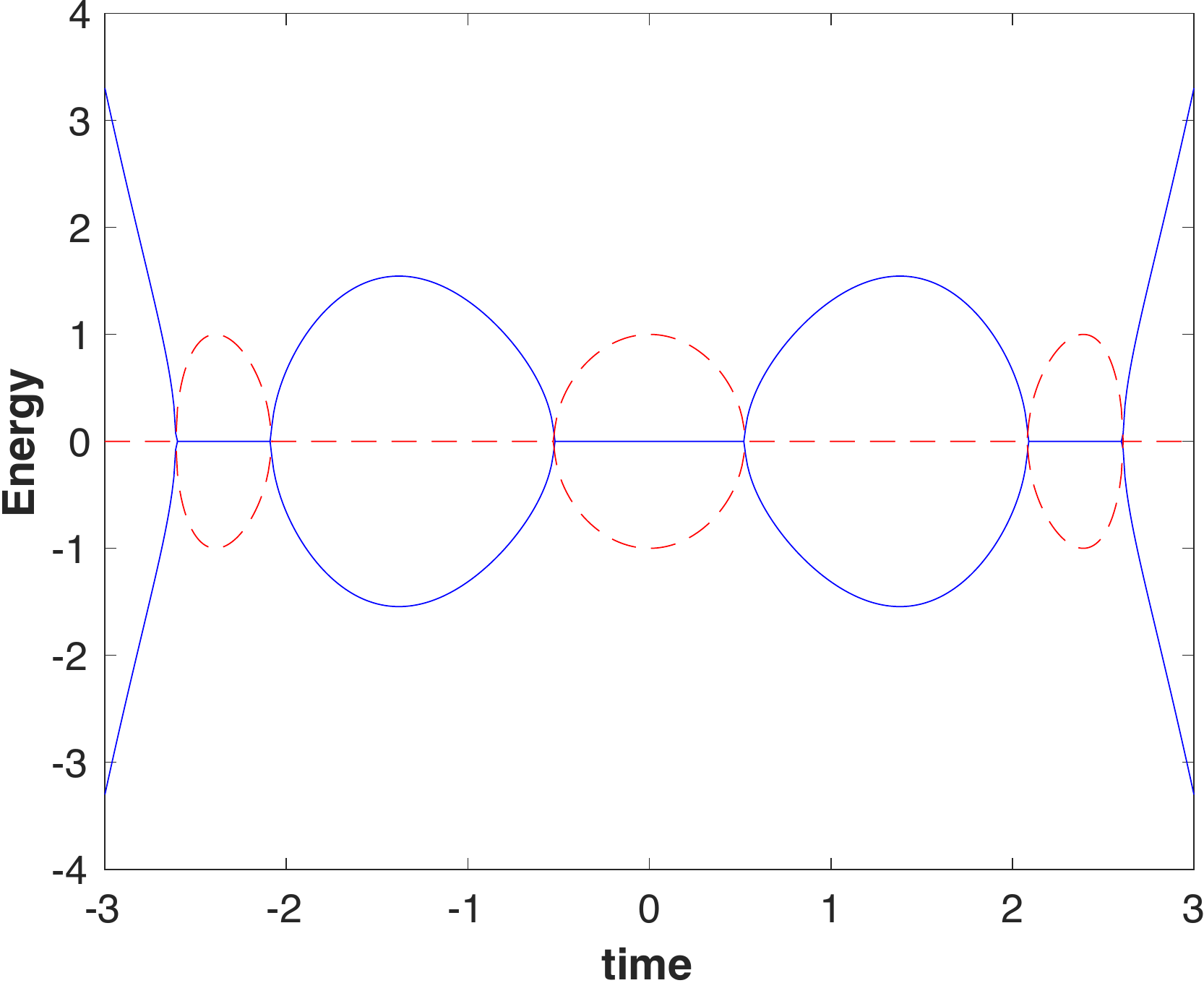}%
}\hfill
\subfloat[$v(t)=-2t+0.35t^3,\Gamma=1$\label{sfig:Fig_1d}]{%
  \includegraphics[width=0.48\columnwidth]{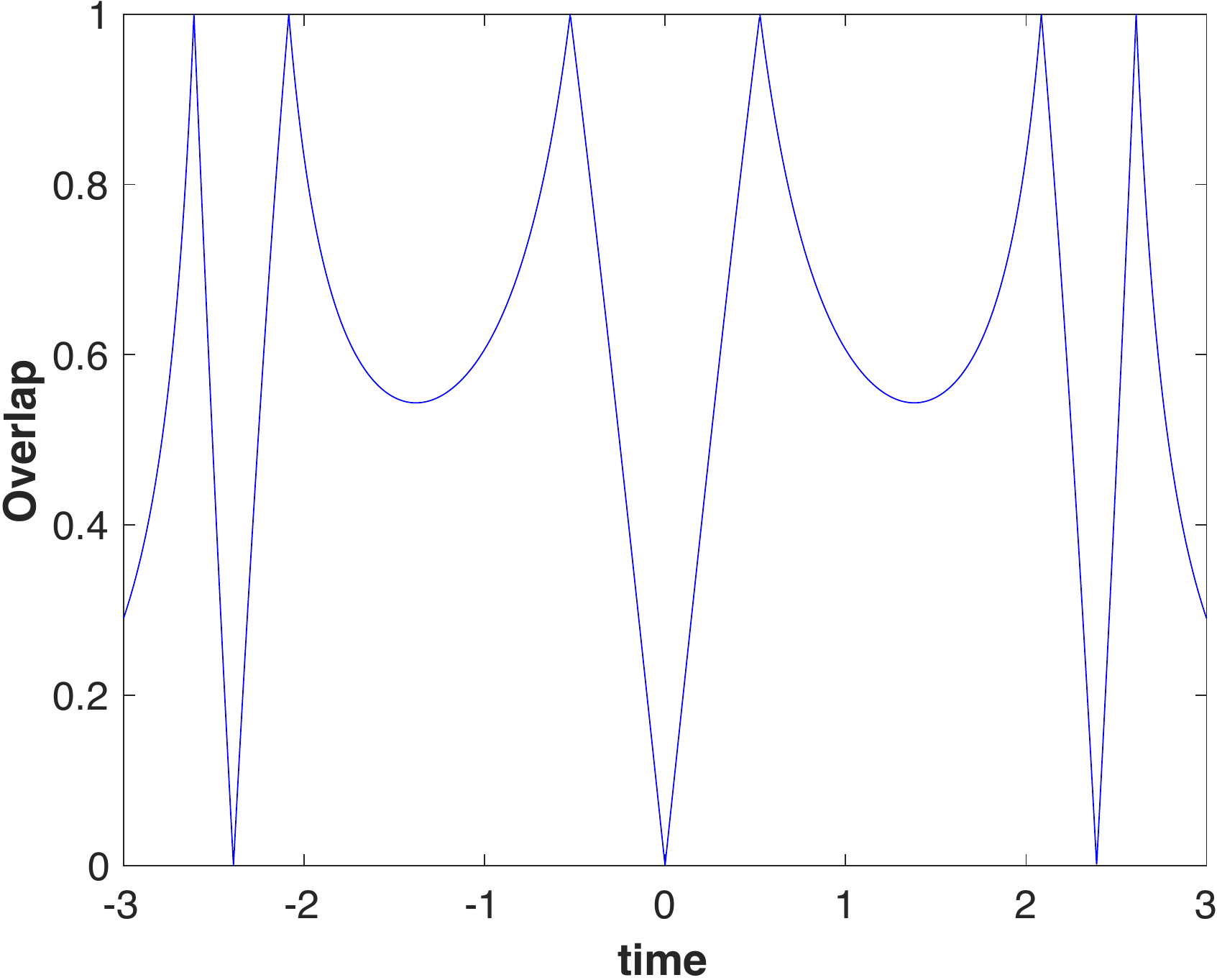}%
}
\caption{Schematic of the energies and overlap of the adiabatic eigenstates for the super-parabolic case with $v(t)=\alpha t +\gamma t^3$ and $\Gamma=\mbox{const}$. The real and imaginary parts of the eigen-energies are shown in blue solid lines and red dash lines respectively.}
\label{fig:NonHermitianModel}
\end{center}
\end{figure}\begin{equation}
    |\phi_{\pm}\rangle=\frac{1}{N_\pm}
    \begin{pmatrix}
    i\Gamma \\
    v+\epsilon_\pm
    \end{pmatrix},
    \langle\chi_\pm|=\frac{1}{N_\pm}\begin{pmatrix}
    i\Gamma & v+\epsilon_\pm
    \end{pmatrix},
\end{equation}
where $N_\pm\equiv \sqrt{2\epsilon_\pm(v+\epsilon_\pm)}$ are the normalization constants that give the inner products $\langle\chi_\pm|\phi_\mp\rangle=0$ and $\langle\chi_\pm|\phi_\pm\rangle=1$. As a result, we obtain the eigen-decomposition of the Hamiltonian, $\hat{H}=\epsilon_+|\phi_+\rangle\langle\chi_+|+\epsilon_-|\phi_-\rangle\langle\chi_-|$, as long as there is no energy degeneracy, i.e., $\epsilon_+\neq\epsilon_-$. The overlap of the instantaneous adiabatic eigenstates $|\phi_+\rangle$ and $|\phi_-\rangle$ has the form
\begin{equation}
   g\equiv \frac{|\langle\phi_-|\phi_+\rangle|}{\sqrt{|\langle\phi_-|\phi_-\rangle\langle\phi_+|\phi_+\rangle|}}=	\left\{\begin{aligned}
 	 \Gamma/|v|,\:\mbox{for}\:|v|\geq\Gamma,\\
  	|v|/\Gamma,\:\mbox{for}\:|v|\leq\Gamma,
	\end{aligned}
	\right.
\end{equation}
which becomes unity at the exceptional points and vanishes when $v$ drops to zero (see Figs.\:\ref{fig:NonHermitianModel2} and \ref{fig:NonHermitianModel}). 

To continue, let us denote the state of the system as $|\psi(t)\rangle\equiv(\psi_1(t),\psi_2(t))^T$, where $\psi_1(t)$ and $\psi_2(t)$ are the wave amplitudes in the diabatic basic. When the conditions $|\psi_1(-\infty)|=0$ and $|\psi_2(-\infty)|=1$ are initially fulfilled, the problem of non-adiabatic transition is to determine the transmission probabilities at $t\rightarrow+\infty$ given by
\begin{equation}
    P_1(t)\equiv \frac{|\psi_1(t)|^2}{|\psi_1(t)|^2+|\psi_2(t)|^2},
    P_2(t)\equiv 1-P_1(t).
\end{equation}
From the Hamiltonian Eq.\:\eqref{NonHermiteHamiltonian}, one obtains the Schr\"{o}dinager equations for the two wave amplitudes $\psi_1$ and $\psi_2$
\begin{subequations}
\begin{align}
    i\dot{\psi}_1&=i\Gamma\psi_2-v(t)\psi_1,\\
    i\dot{\psi}_2&=i\Gamma\psi_1+v(t)\psi_2,
\end{align}
\end{subequations}
from which one immediately obtains the second-order differential equations that $\psi_1$ and $\psi_2$ obey
\begin{subequations}      
\begin{gather}
    \ddot{\psi}_1+[v^2(t)-\Gamma^2-i\dot{v}(t)]\psi_1=0,\label{dd1}\\
    \ddot{\psi}_2+[v^2(t)-\Gamma^2+i\dot{v}(t)]\psi_2=0\label{dd2}.
\end{gather}
\end{subequations}
For unitary evolutions, $|\psi_2(t)|^2$ can be calculated from $|\psi_1(t)|^2$ due to the conservation of total population, i.e., $|\psi_1(t)|^2+|\psi_2(t)|^2=\mbox{const}$. However, for non-unitary evolutions, the total population is in general not conserved, and hence $|\psi_2(t)|^2$ has to be calculated separately even when $|\psi_1(t)|^2$ is known. Fortunately, for our non-Hermitian two level model, we have
\begin{equation*}
\frac{d}{dt}|\psi_1|^2=\frac{d}{dt}|\psi_2|^2=\Gamma(\psi_1^*\psi_2+\psi_2^*\psi_1),
\end{equation*}
which implies that the difference between the level populations, $|\psi_2|^2-|\psi_1|^2$, is still a constant. For the special case that the separation of diabatic energies varies linearly with time, i.e., $v(t)=\alpha t$ and $\Gamma=\mbox{const}$, Eqs.\:\eqref{dd1} - \eqref{dd2} describe the non-Hermitian generalization of the Landau-Zener model, which has the following exact solutions for the final level populations: $|\psi_1(\infty)|^2=e^{\pi\Gamma^2/\alpha}-1$ and $|\psi_2(\infty)|^2=e^{\pi\Gamma^2/\alpha}$, provided that the system is initially subjected to the constraints $\psi_1(-\infty)=0$ and $|\psi_2(-\infty)|=1$. In general, when the linear separation of the diabatic energies is modified by an additional term $c_nt^n$, i.e., $v(t)\equiv \alpha t+c_nt^n$ and $\Gamma\equiv\mbox{const}$, $\psi_1$ and $\psi_2$ are governed by
\begin{align}\label{SecondODE}
    \ddot{\psi}_1&+(c_n^2t^{2n}+2\alpha c_nt^{n+1}-inc_nt^{n-1}+\alpha^2t^2-\Gamma^2-i\alpha)\psi_1=0,\nonumber\\
    \ddot{\psi}_2&+(c_n^2t^{2n}+2\alpha c_nt^{n+1}+inc_nt^{n-1}+\alpha^2t^2-\Gamma^2+i\alpha)\psi_2=0.\nonumber
\end{align}
Clearly, unlike the conventional Hermitian models associated with $n$ level-crossing points, there are in general $2n$ exceptional points for the non-Hermitian systems. 

For the parabolic case with $c_2\equiv\beta\neq 0$, we obtain $\ddot{\psi}_1+(\beta^2t^4+2\alpha\beta t^3+\alpha^2t^2-2i\beta t-\Gamma^2-i\alpha)\psi_1=0$ and $\ddot{\psi}_2+(\beta^2t^4+2\alpha\beta t^3+\alpha^2t^2+2i\beta t-\Gamma^2+i\alpha)\psi_2=0$. After a change of variable $\tau\equiv t+\frac{\alpha}{2\beta}$, the equation that governs $\psi_1$ becomes\begin{figure}[tbp]
\begin{center}
\subfloat[$v(t)=\alpha t+\beta t^2$\label{sfig:Fig_5a}]{%
  \includegraphics[width=0.48\columnwidth]{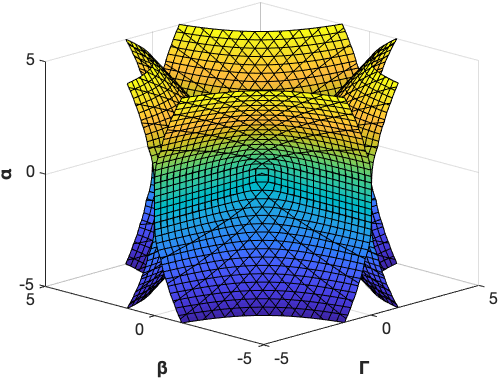}%
}\hfill
\subfloat[$v(t)=\alpha t+\gamma t^3$\label{sfig:Fig_5b}]{%
  \includegraphics[width=0.48\columnwidth]{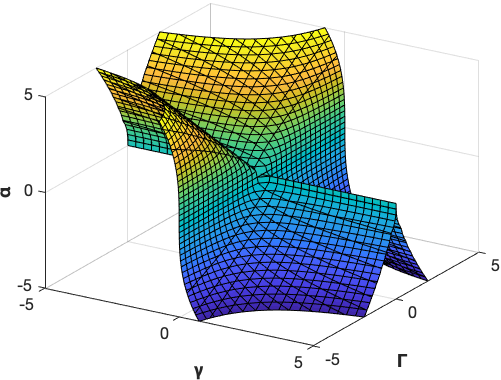}%
}
\caption{Schematic of the critical surfaces in the parameter space for the parabolic and super-parabolic non-Hermitian two-level models. In Fig.\:\ref{sfig:Fig_5a}, the critical surface is determined by $\alpha^4=16\Gamma^2\beta^2$. There are four exceptional points for the space outside the critical surface; whereas there are two exceptional points for the space within the critical surface. On the critical surface, there are three exceptional points. In Fig.\:\ref{sfig:Fig_5b}, the critical surface is determined by $4\alpha^3=-27\gamma\Gamma^2$. There are six exceptional points for the space beneath the critical surface; whereas there are two exceptional points for the space above the critical surface. On the critical surface, there are four exceptional points.}
\end{center}
\end{figure}
\begin{equation}\label{TriconfluentHeun}
    \frac{d^2\psi_1}{d\tau^2}+\left[\left(\beta\tau^2-\frac{\alpha^2}{4\beta}\right)^2-\Gamma^2-2i\beta\tau\right]\psi_1=0,
\end{equation}
where $\psi_2$ satisfies a similar equation with $-2i\beta\tau$ replaced by $2i\beta\tau$. For this system, there are at most four exceptional points located at $\tau=\pm\frac{1}{2|\beta|}\sqrt{\alpha^2\pm 4\Gamma\beta^2}$. For $\alpha^4>16\Gamma^2\beta^2$, there are four exceptional points; for $\alpha^4<16\Gamma^2\beta^2$, by contrast, there are only two exceptional points; whereas for $\alpha^4=16\Gamma^2\beta^2$, there are three exceptional points. Hence, $\alpha^4=16\Gamma^2\beta^2$ defines a critical surface in the parameter space (see Fig.\:\ref{sfig:Fig_5a}). After the transformation $z\equiv \sqrt{2i\beta/3}\tau$, Eq.\:\eqref{TriconfluentHeun} becomes the second canonical form of the tri-confluent Heun equation
\begin{equation}
    \frac{d^2\psi_1}{dz^2}+\left(\mu-\frac{\xi^2}{4}+\nu z-\frac{3}{2}\xi z^2-\frac{9}{4}z^4\right)\psi_1=0,
\end{equation}
where $\mu\equiv -\Gamma^2$, $\nu\equiv -\sqrt{6i\beta}$ and $\xi\equiv-i\alpha^2/(2\beta)$. 

For the super-parabolic case with $c_3\equiv\gamma\neq 0$, we obtain $\ddot{\psi}_1+(\gamma^2t^6+2\alpha\gamma t^4+(\alpha^2-3i\gamma)t^2-\Gamma^2-i\alpha)\psi_1=0$ and $\ddot{\psi}_2+(\gamma^2t^6+2\alpha\gamma t^4+(\alpha^2+3i\gamma)t^2-\Gamma^2+i\alpha)\psi_2=0$. There are at most six exceptional points determined by the cubic equations $\gamma t^3+\alpha t\pm\Gamma=0$. Without loss of generality, we may assume that $\gamma>0$. For $4\alpha^3+27\gamma\Gamma^2<0$, there are six exceptional points; for $4\alpha^3+27\gamma\Gamma^2>0$, by contrast, there are only two exceptional points; whereas for $4\alpha^3+27\gamma\Gamma^2=0$, there are four exceptional points. Hence, $4\alpha^3+27\gamma\Gamma^2=0$ defines a critical surface in the parameter space (see Fig.\:\ref{sfig:Fig_5b}). After the transformations $\tau\equiv t^2$ and  $U_1\equiv \tau^{1/4}\psi_1$, the equation that governs $\psi_1$ becomes
\begin{equation}
    \frac{d^2U_1}{d\tau^2}+\left(\frac{3}{16\tau^2}-\frac{\Gamma^2+i\alpha}{4\tau}+\frac{\alpha^2-3i\gamma}{4}+\frac{\alpha\gamma \tau}{2} +\frac{\gamma^2\tau^2}{4}\right)U_1=0.
\end{equation}
After another change of variable $\xi\equiv\sqrt{-i\gamma/2}\tau$, it becomes the second canonical form of the bi-confluent Heun equation
\begin{equation}
    \frac{d^2U_1}{d\xi^2}+\left(\frac{1-\mu^2}{4\xi^2}-\frac{\eta}{2\xi}+\lambda-\frac{\nu^2}{4}-\nu\xi-\xi^2\right)U_1=0,
\end{equation}
where $\mu=-\frac{1}{2}$, $\nu=\alpha\sqrt{\frac{-2i}{\gamma}}$, $\lambda=\frac{3}{2}$ and $\eta=-\frac{\nu}{2}(1+\frac{\Gamma^2}{i\alpha})$.

\section{Analytical Approximations to the transmission probabilities}\label{III}
To visualize and analyze the non-Hermitian dynamics, one may introduce four real variables, i.e., $S_0\equiv|\psi_1|^2+|\psi_2|^2$, $S_1\equiv \psi_2^*\psi_1+\psi_1^*\psi_2$, $S_2\equiv -i(\psi_2^*\psi_1-\psi_1^*\psi_2)$ and $S_3\equiv|\psi_2|^2-|\psi_1|^2$, which obey $S_0^2-S_1^2-S_2^2=S_3^2=\mbox{const}$. As one may write $S_1\equiv(S_0^2-S_3^2)^{1/2}\cos\Theta$ and $S_2\equiv(S_0^2-S_3^2)^{1/2}\sin\Theta$ with $\Theta\equiv\arg{\psi_1}-\arg{\psi_2}$ being the relative phase between the two wave amplitudes $\psi_1$ and $\psi_2$, $S_1$ and $S_2$ are related to the total level population $S_0$ and the relative phase $\Theta$. For the case when the system is initially in the instantaneous eigenstates, we obtain $S_3=\pm 1$. Hence, both $|\psi_2|^2\equiv\frac{1}{2}(S_0\pm 1)$ and $|\psi_1|^2\equiv\frac{1}{2}(S_0\mp 1)$ can be determined from the total level population $S_0$.
\begin{figure}[tbp]
\begin{center}
\subfloat[$v(t)=t+t^3,\Gamma=1$\label{sfig:Fig_2a}]{%
  \includegraphics[width=0.5\columnwidth]{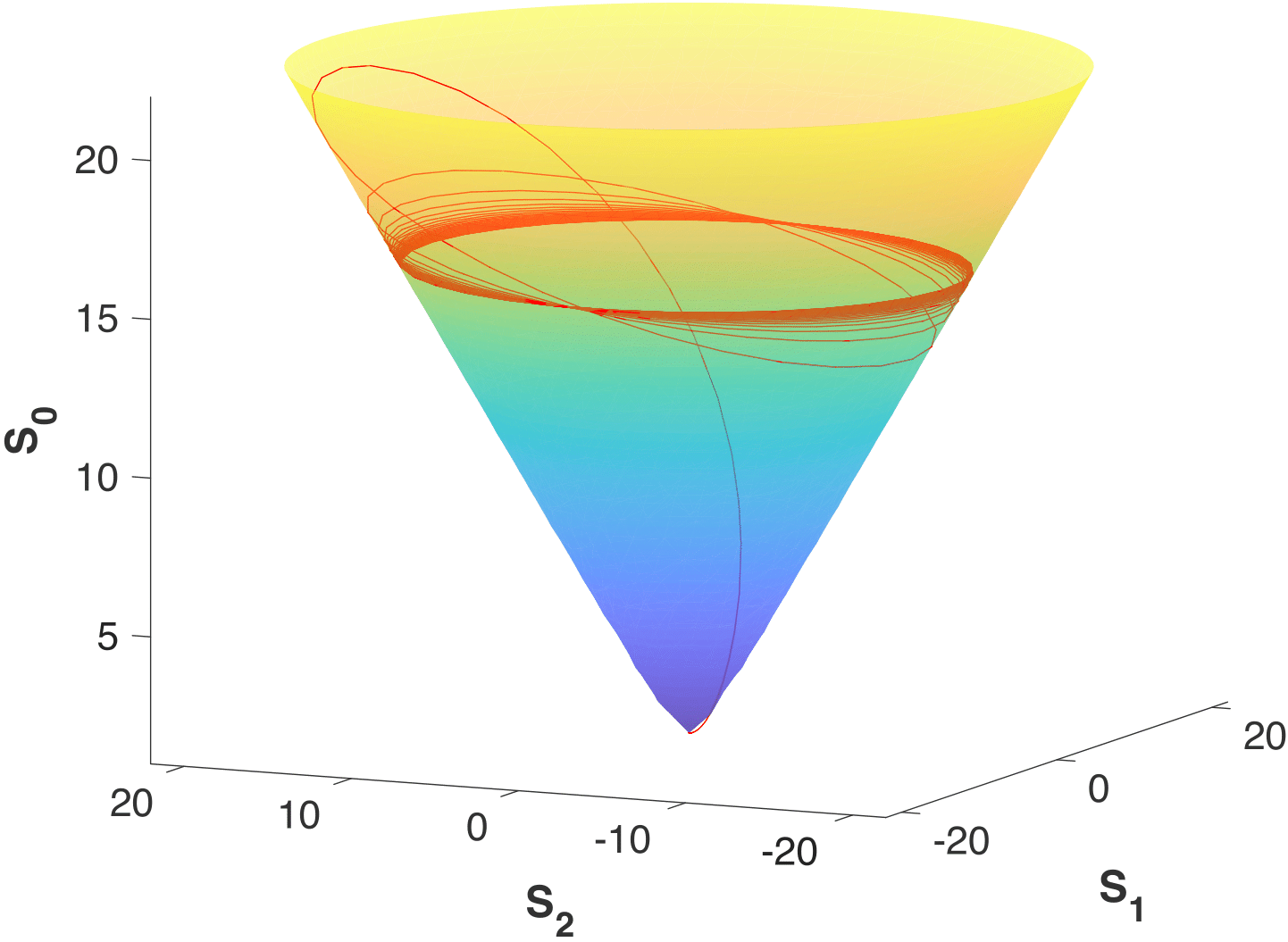}%
}\hfill
\subfloat[$v(t)=t+t^3,\Gamma=1$\label{sfig:Fig_2b}]{%
  \includegraphics[width=0.48\columnwidth]{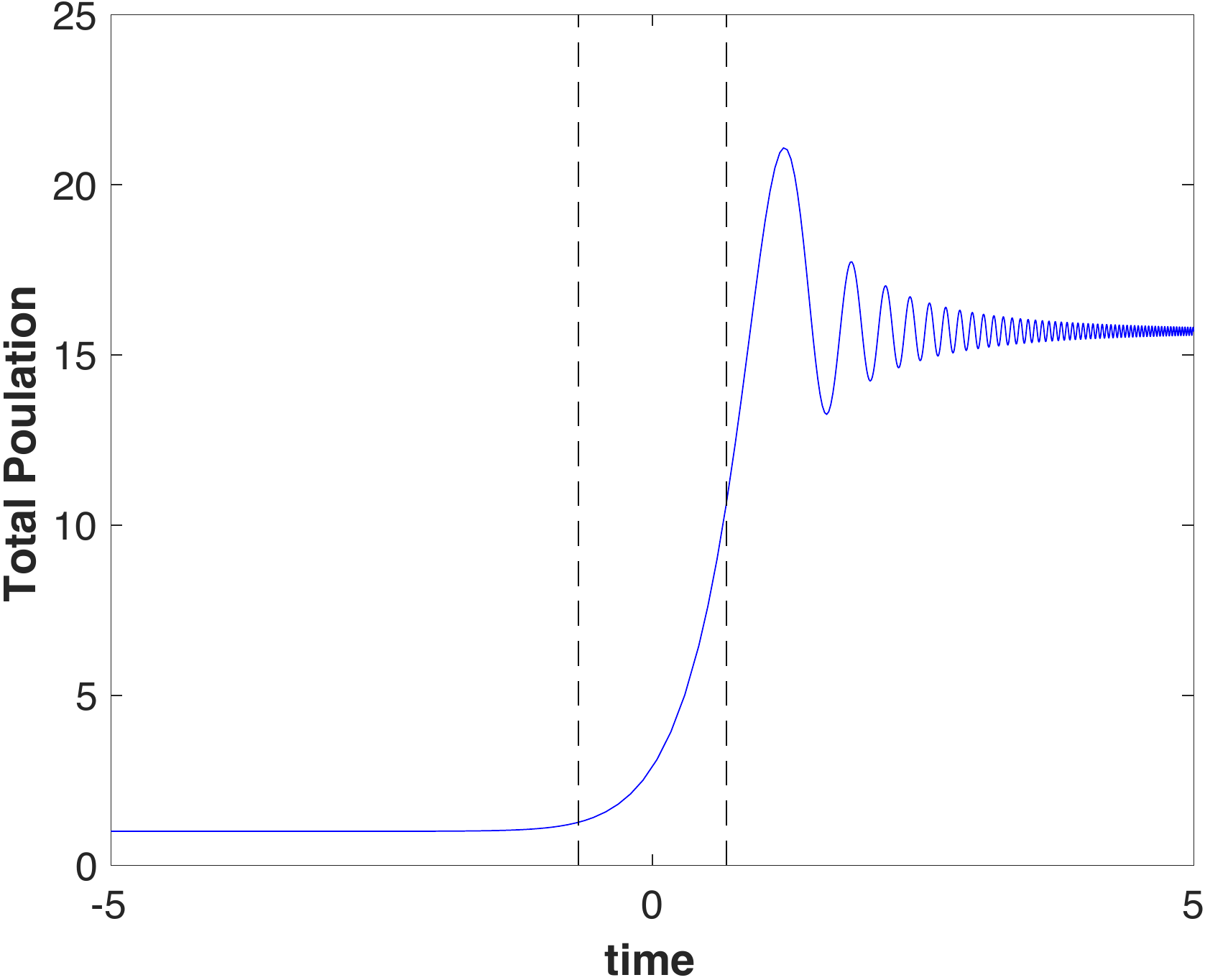}%
}\hfill
\subfloat[$v(t)=-2t+0.35t^3,\Gamma=1$\label{sfig:Fig_2c}]{%
  \includegraphics[width=0.48\columnwidth]{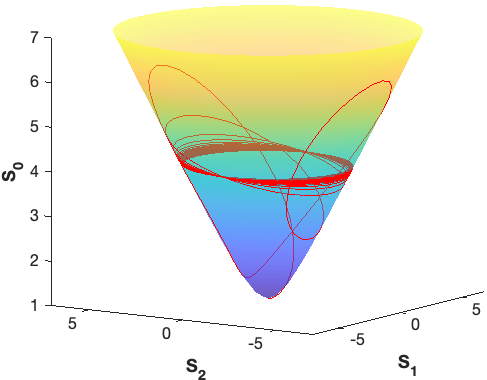}%
}\hfill
\subfloat[$v(t)=-2t+0.35t^3,\Gamma=1$\label{sfig:Fig_2d}]{%
  \includegraphics[width=0.48\columnwidth]{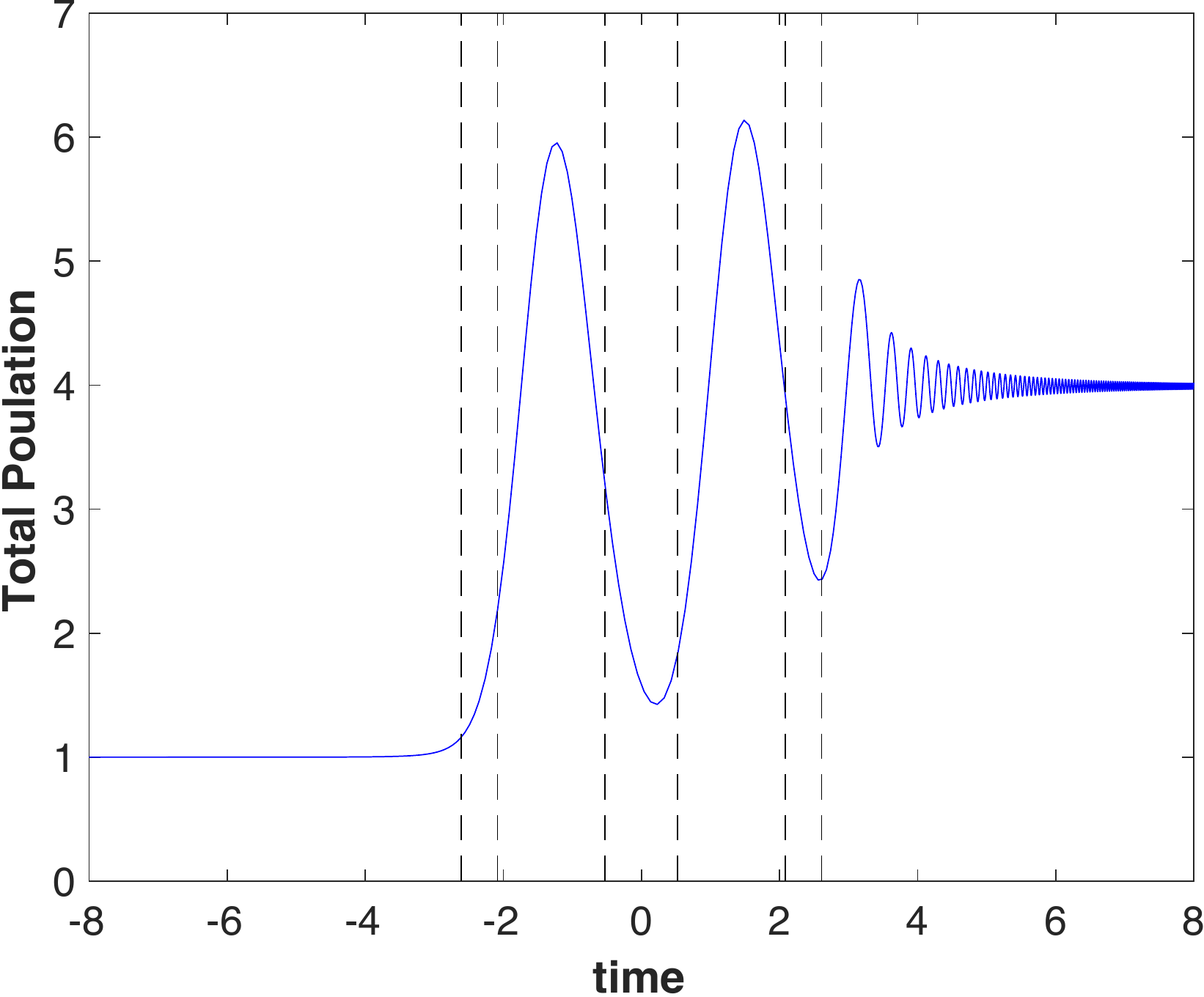}%
}
\caption{Schematic of the non-Hermitian two-level dynamics generated by the Hamiltonian \eqref{NonHermiteHamiltonian}, where the separation of diabatic energies $2v$ varies in time as $v(t)=\alpha t+\gamma t^3$. In Fig.\:\ref{sfig:Fig_2a} and Fig.\:\ref{sfig:Fig_2c}, the dynamics in terms of the variables $S_0$, $S_1$ and $S_2$ are shown on an unit hyperboloid of two sheets in red solid lines. In Fig.\:\ref{sfig:Fig_2b} and Fig.\:\ref{sfig:Fig_2d}, the total population $S_0=|\psi_1|^2+|\psi_2|^2$ are shown in blue solid lines, and the exceptional points are shown as dashed lines.}
\label{fig:NonHermitianSpin}
\end{center}
\end{figure} 

The non-Hermitian two-level dynamics may be visualized on a hyperboloid of two sheets with $S_0$ being the horizontal direction (see Figs.\:\ref{sfig:Fig_2a} and \ref{sfig:Fig_2c}), which is described by the following set of differential equations\begin{subequations}
\begin{align}
    \dot{S}_0&=2\Gamma S_1,\label{5a}\\
    \dot{S}_1&=2\Gamma S_0-2v(t)S_2,\label{5b}\\
    \dot{S}_2&=2v(t)S_1.\label{5c}
\end{align}
\end{subequations}
\begin{figure}[tbp]
\begin{center}
\subfloat[$v(t)=t+t^3,\Gamma=1$\label{sfig:Fig_3a}]{%
  \includegraphics[width=0.48\columnwidth]{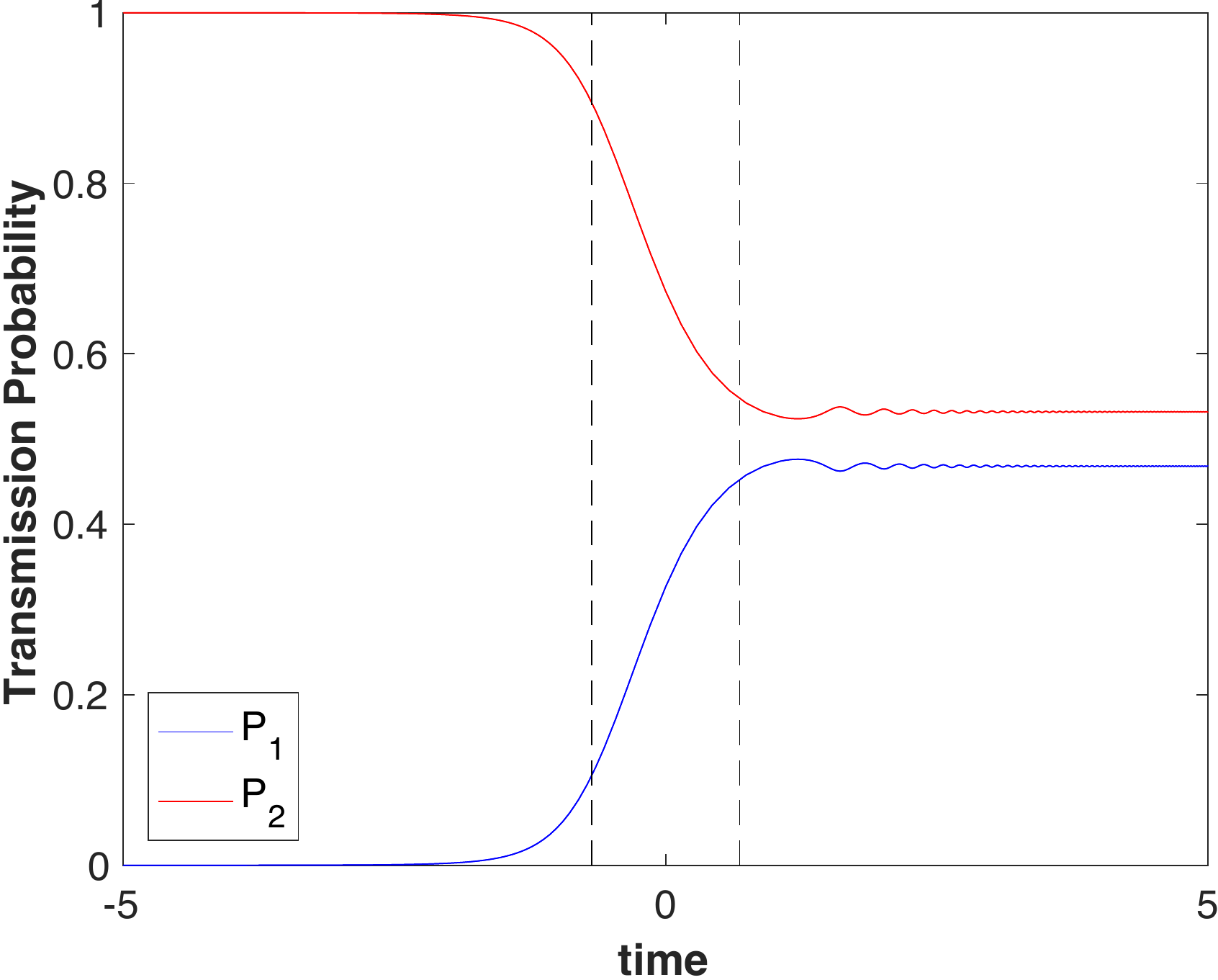}%
}\hfill
\subfloat[$v(t)=t+t^3,\Gamma=1$\label{sfig:Fig_3c}]{%
  \includegraphics[width=0.48\columnwidth]{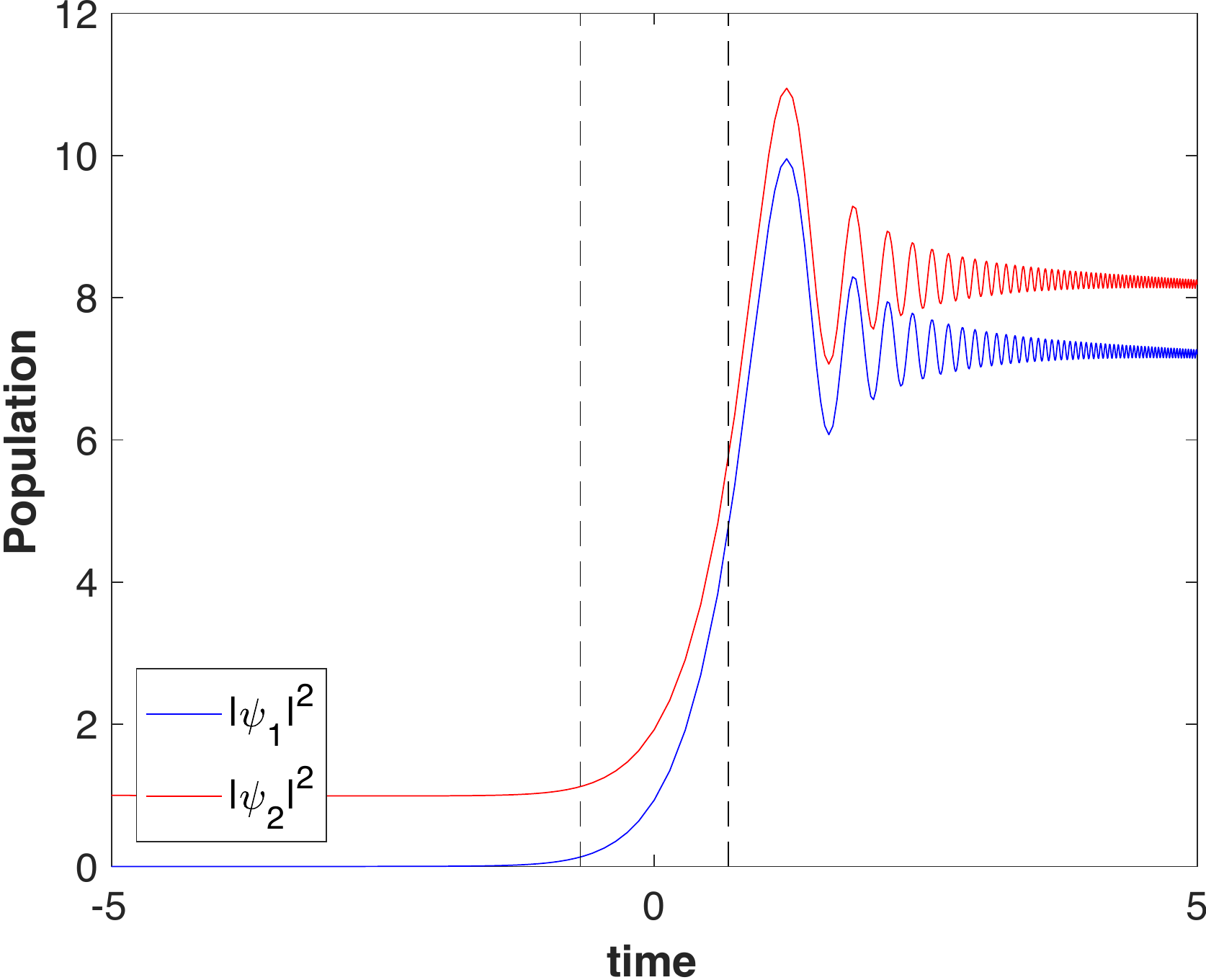}%
}\hfill
\subfloat[$v(t)=-2t+0.35t^3,\Gamma=1$\label{sfig:Fig_3b}]{%
  \includegraphics[width=0.48\columnwidth]{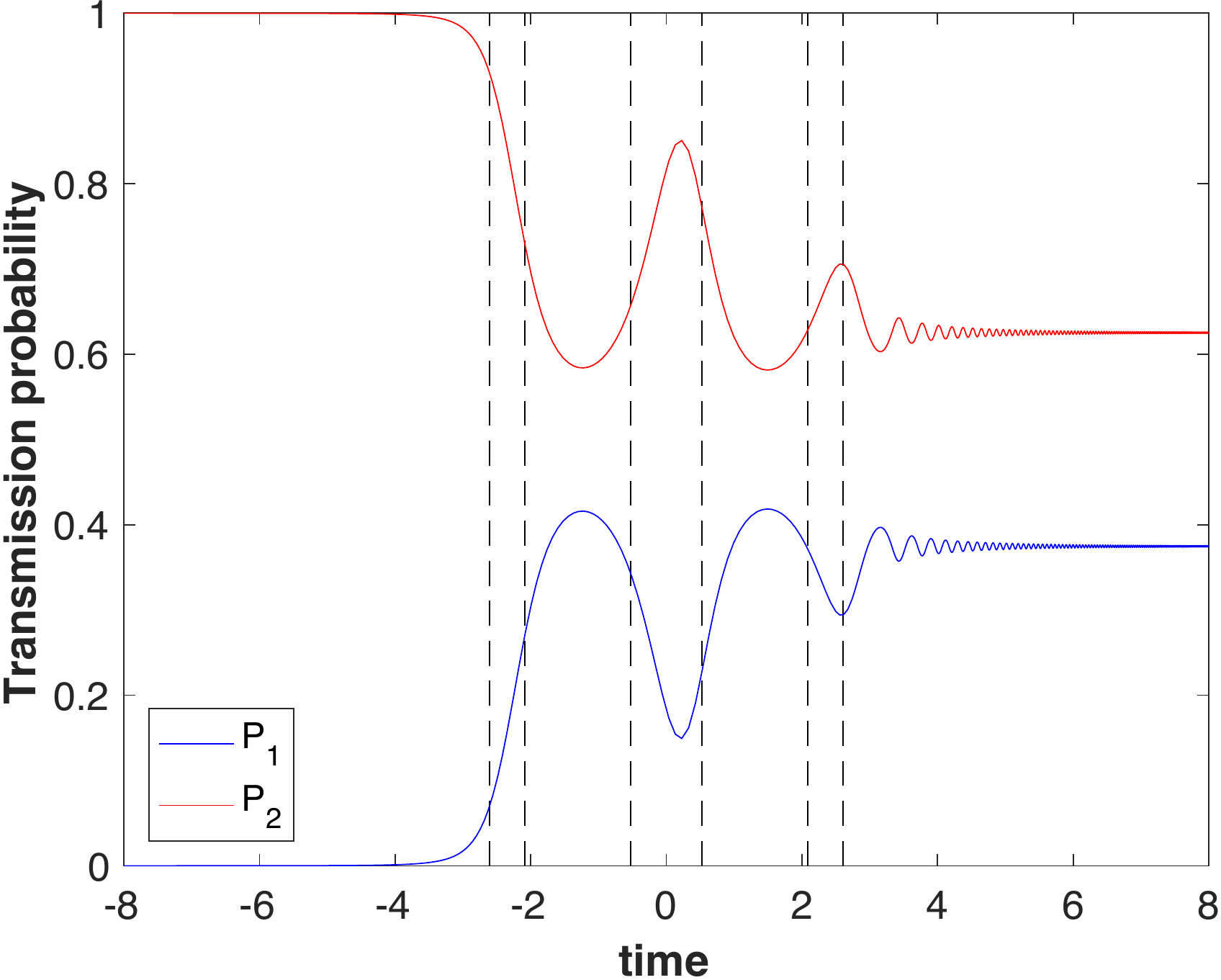}%
}\hfill
\subfloat[$v(t)=-2t+0.35t^3,\Gamma=1$\label{sfig:Fig_3d}]{%
  \includegraphics[width=0.48\columnwidth]{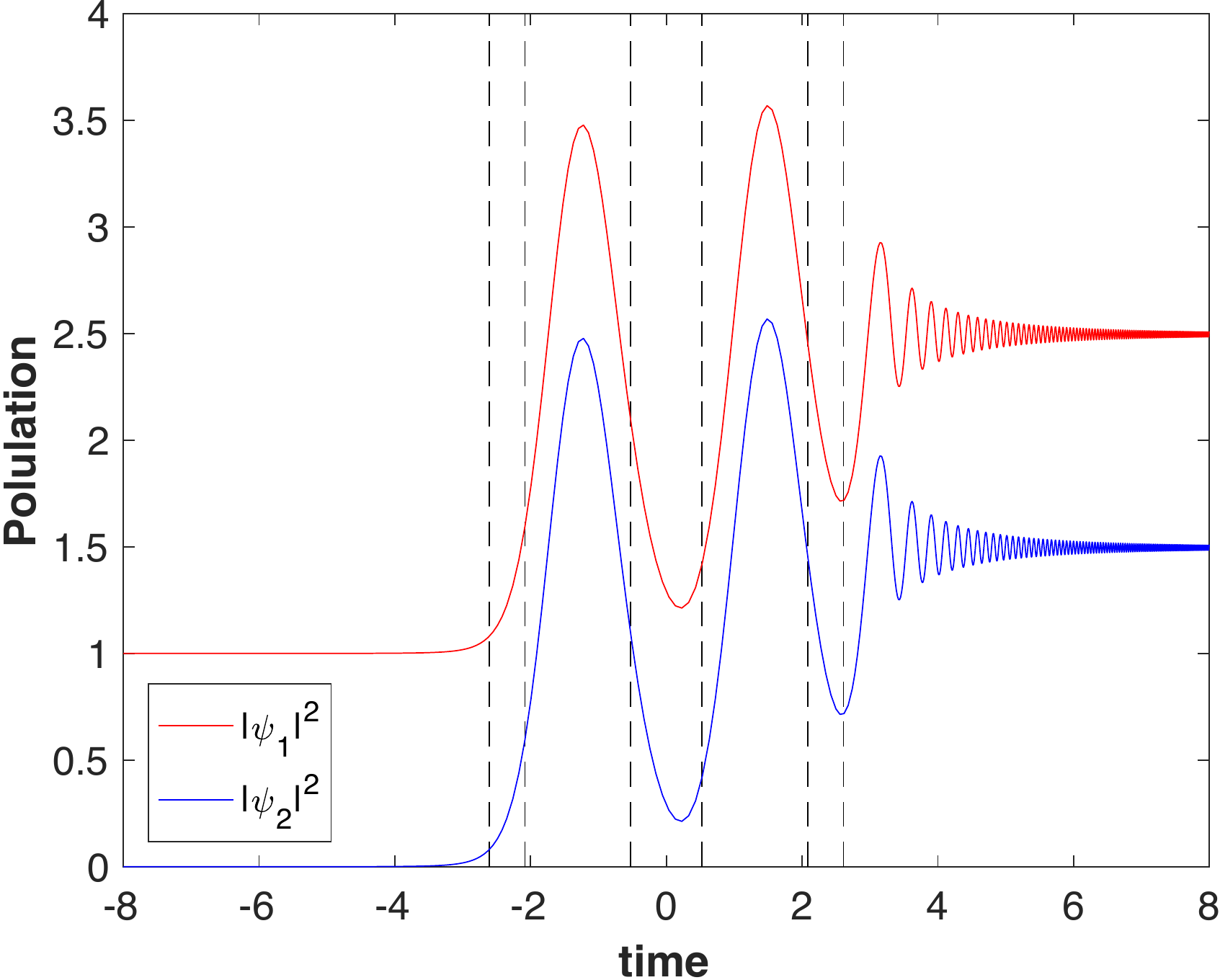}%
}
\caption{Schematic of the transition probabilities and the level populations as functions of time. In Figs.\:\ref{sfig:Fig_3a} and \ref{sfig:Fig_3b}, the transmission probabilities given by $P_1\equiv |\psi_1|^2/(|\psi_1|^2+|\psi_2|^2)$ and $P_2\equiv |\psi_2|^2/(|\psi_1|^2+|\psi_2|^2)$ are shown as blue and red solid lines respectively. In Fig.\:\ref{sfig:Fig_3a}, there are two exceptional points located at $t=\pm 0.68233$, which are shown as dashed lines. In Fig.\:\ref{sfig:Fig_3b}, there are six exceptional points located at $t=\pm 3.38762$, $\pm2.87408$ and $\pm0.51354$. In Figs.\:\ref{sfig:Fig_3c} and \ref{sfig:Fig_3d}, the level populations $|\psi_1|^2$ and $|\psi_2|^2$ are shown as blue and red solid lines respectively.}
\label{fig:TransitionProbablity}
\end{center}
\end{figure}

When the system is initially ($t=t_i$) in one of the instantaneous eigenstates with $|v|\gg \Gamma$, the coupling between the two states is weak. Thus, the system is expected to follow the initial instantaneous eigenstate, until the first exceptional point is reached. For the case when there are only two exceptional points at $t=t_1$ and $t=t_2$ ($t_1<t_2$), the two wave amplitudes $\psi_1$ and $\psi_2$ as well as the total population can be regarded as constants for $t\leq t_1$. After the first exceptional point is reached, the total population grows exponentially in the region of broken $\mathcal{P}\mathcal{T}$ symmetry, i.e., $t_1\leq t\leq t_2$, which is also demonstrated by the fast escaping trajectory on the hyperboloid (see Figs.\:\ref{sfig:Fig_2a} and \ref{sfig:Fig_2b}). As the coupling between the two states is strong compared to $|v|$ in the region of broken $\mathcal{P}\mathcal{T}$ symmetry, one may neglect the term $-2vS_2$ in Eq.\:\eqref{5b}, which yields $\ddot{S}_0=4\Gamma^2S_0$. The total population $S_0$ as well as the other two variables $S_1$ and $S_2$ for $t_1\leq t\leq t_2$ can be approximated by
\begin{figure}[tbp]
\begin{center}
\subfloat[$v(t)=t+t^2,\Gamma=1$\label{sfig:Fig_6a}]{%
  \includegraphics[width=0.49\columnwidth]{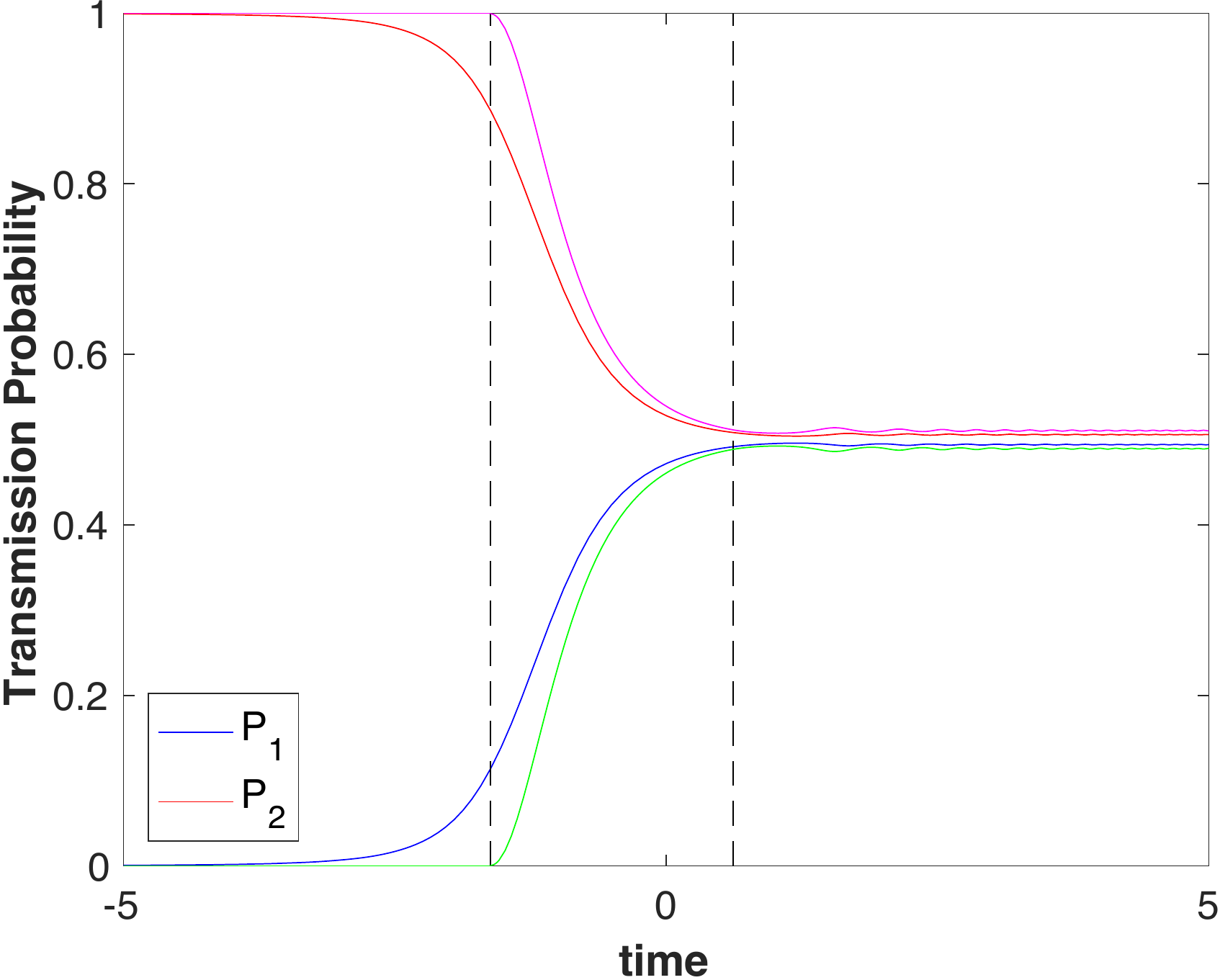}%
}\hfill
\subfloat[$v(t)=t+t^2,\Gamma=1$\label{sfig:Fig_6b}]{%
  \includegraphics[width=0.49\columnwidth]{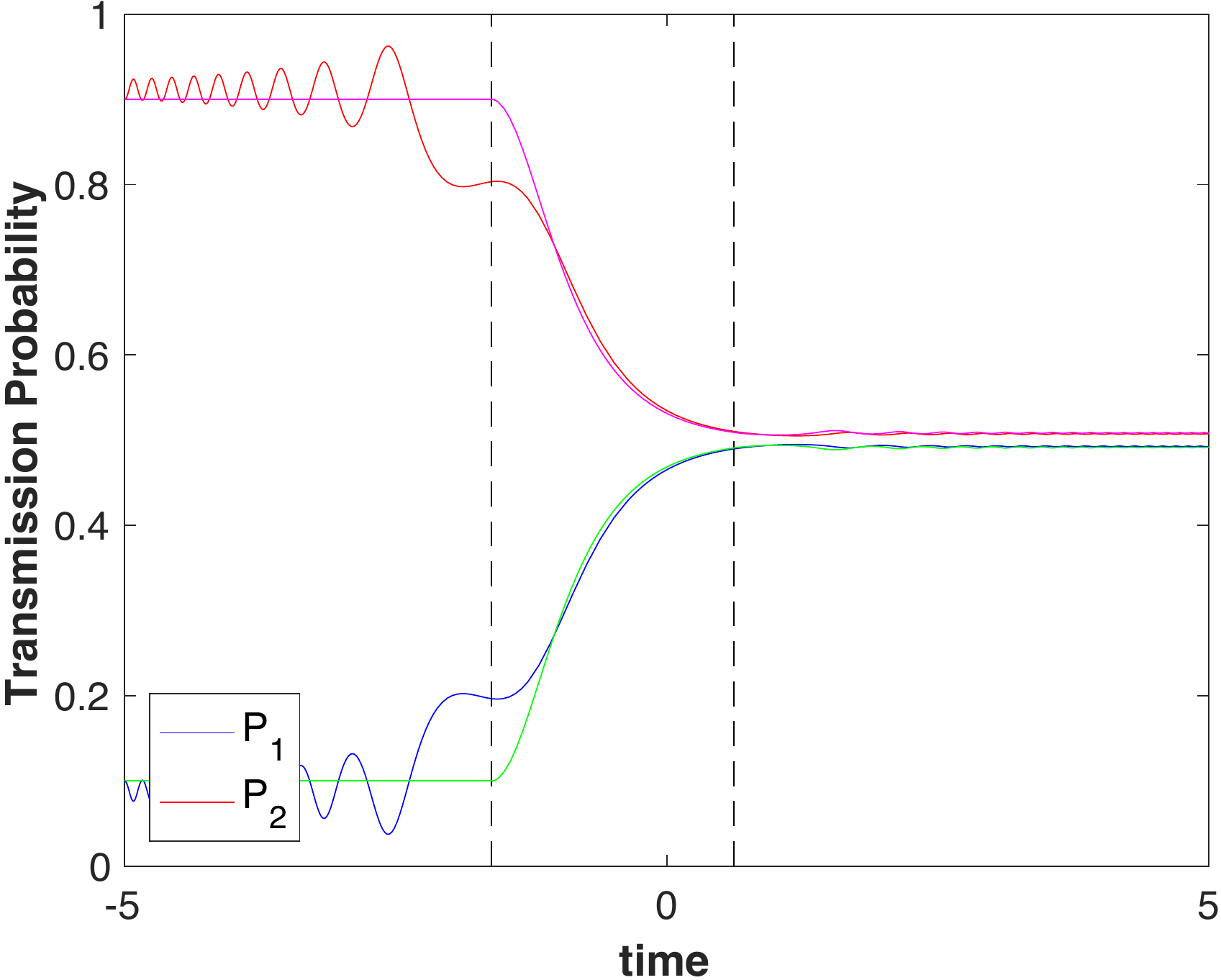}%
}\hfill
\subfloat[$v(t)=t+t^3,\Gamma=1$\label{sfig:Fig_4a}]{%
  \includegraphics[width=0.49\columnwidth]{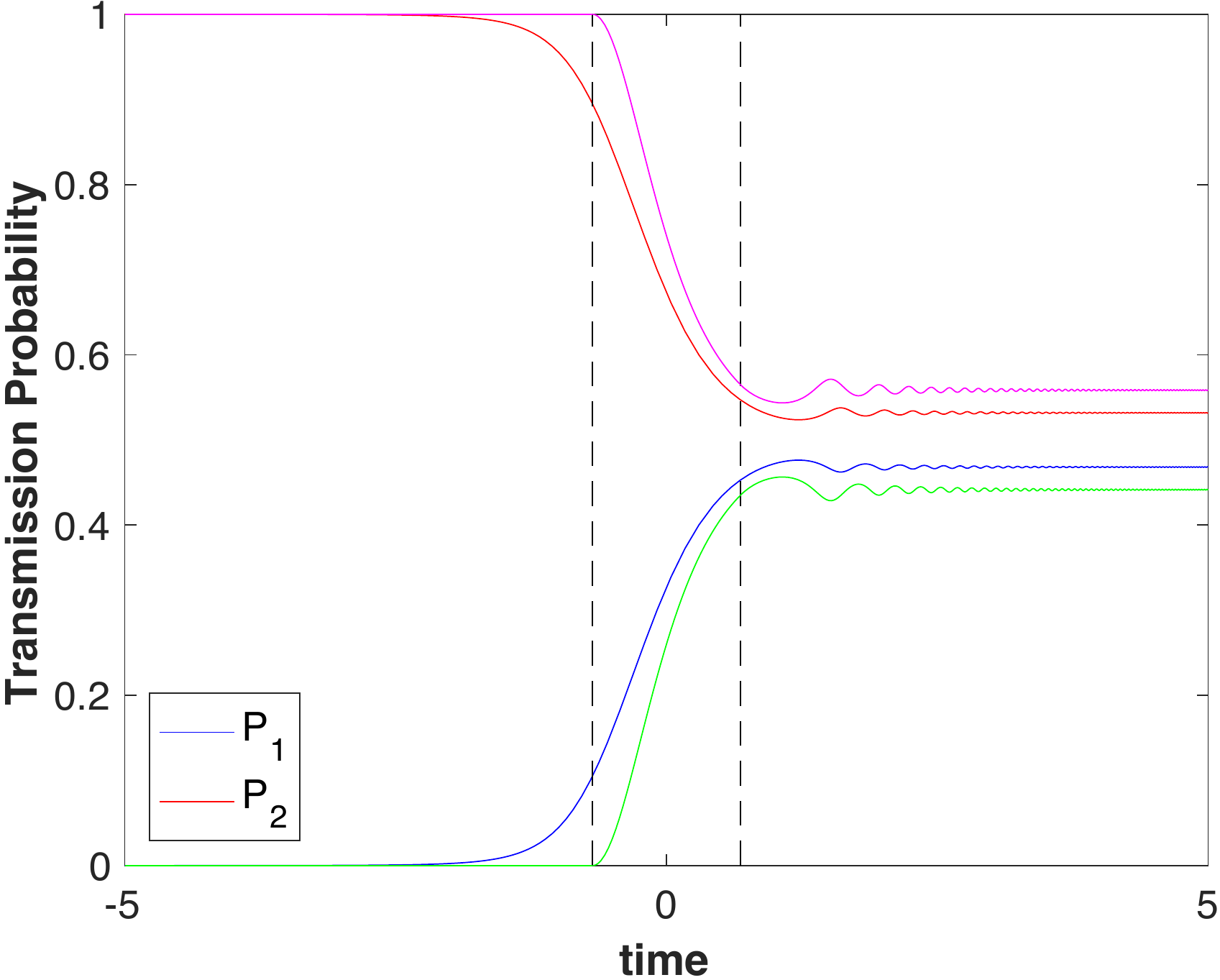}%
}\hfill
\subfloat[$v(t)=t+t^3,\Gamma=1$\label{sfig:Fig_4b}]{%
  \includegraphics[width=0.49\columnwidth]{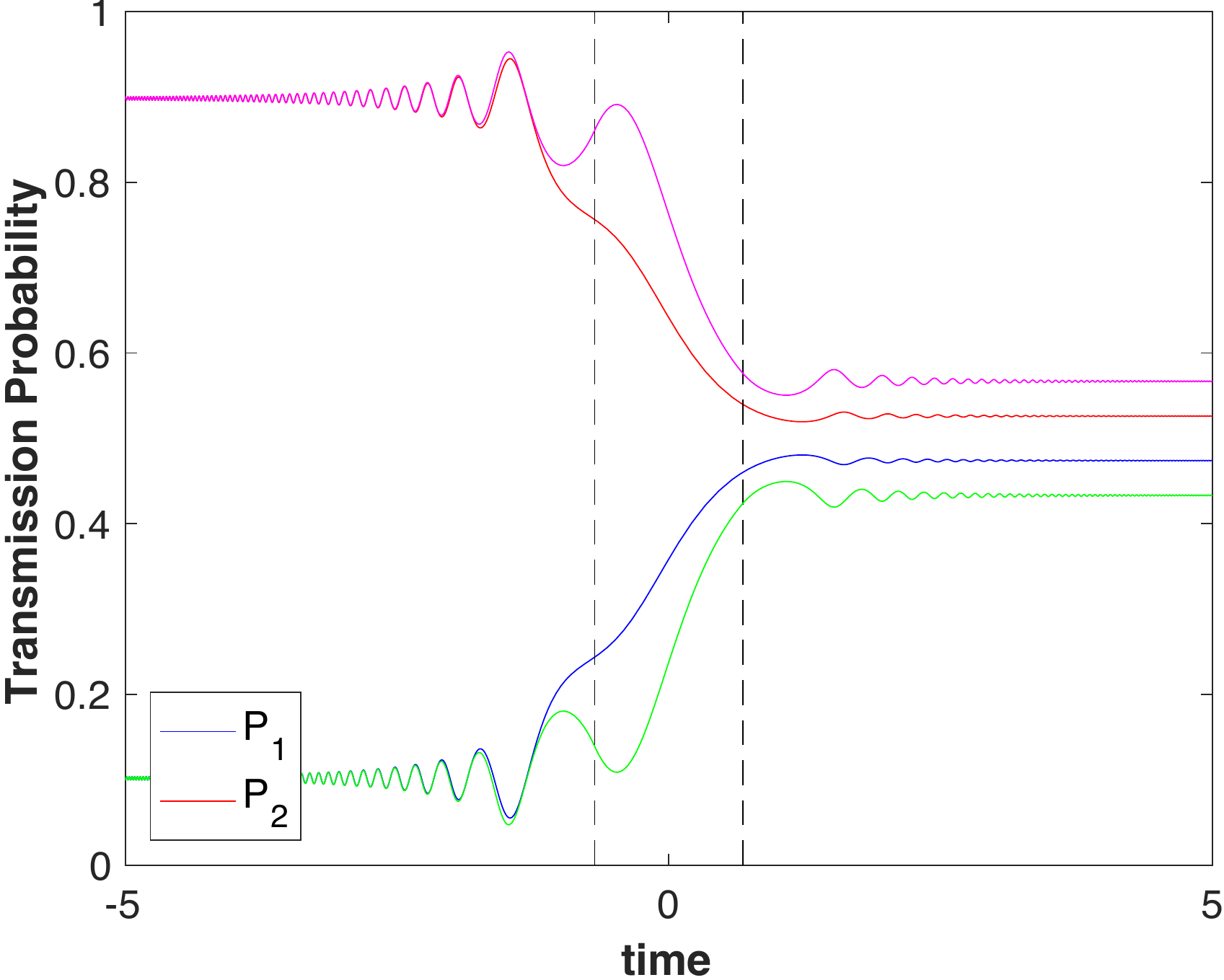}%
}
\caption{Schematic of the transmission probabilities $P_1\equiv|\psi_1|^2/\sum_{i=1}^2|\psi_i|^2$ and $P_2\equiv 1-P_1$ for the parabolic case ($\alpha=\beta=\Gamma=1$) and the super-parabolic case ($\alpha=\gamma=\Gamma=1$). For the parabolic case, there are two exceptional points located at $t=-1.61803$ and $0.61803$ (dashed lines); for the super-parabolic case, there are two exceptional points located at $t=\pm 0.68233$ (dashed lines). In Figs.\:\ref{sfig:Fig_6a} and \ref{sfig:Fig_4a}, the initial state is one of the instantaneous eigenstate; in Figs.\:\ref{sfig:Fig_6b} and \ref{sfig:Fig_4b}, the initial state is a randomly selected state ($P_1=0.1$). The numerical results are shown in red and blue lines; the approximate solutions are shown in magenta and green lines.}
\label{fig:tranS0beta}
\end{center}
\end{figure}
\begin{align}
    S_0(t)&\approx S_0(t_1)\cosh(2\Gamma(t-t_1))+S_1(t_1)\sinh(2\Gamma(t-t_1)),\nonumber\\
    S_1(t)&\approx S_0(t_1)\sinh(2\Gamma(t-t_1))+S_1(t_1)\cosh(2\Gamma(t-t_1)),\nonumber\\
    S_2(t)&\approx S_2(t_1)+2S_0(t_1)\int_{t_1}^tv(s)\sinh(2\Gamma(s-t_1))ds\nonumber\\
    &+2S_1(t_1)\int_{t_1}^tv(s)\cosh(2\Gamma(s-t_1))ds,\label{6}
\end{align}
where $S_k(t_1)\equiv S_k(t_i)$. After leaving the last exceptional point at $t=t_2$, the total population approaches to its stationary value in the region of unbroken $\mathcal{P}\mathcal{T}$ symmetry (see Fig.\:\ref{sfig:Fig_2b}), which may be understood by neglecting the term $2\Gamma S_0$ in Eq.\:\eqref{5b}. Defining $\Phi(t)\equiv 2\int_{t_2}^tv(t)dt$, one obtains
\begin{align}
    S_1(t)&\approx S_1(t_2)\cos\Phi(t)-S_2(t_2)\sin\Phi(t),\nonumber\\
    S_2(t)&\approx S_1(t_2)\sin\Phi(t)+S_2(t_2)\cos\Phi(t),\nonumber\\
    S_0(t)&\approx S_0(t_2)+2\Gamma S_1(t_2)\int_{t_2}^t\cos\Phi(s)ds\nonumber\\
    &-2\Gamma S_2(t_2)\int_{t_2}^t\sin\Phi(s)ds),\label{7}
\end{align}
where $S_k(t_2)$ are evaluated using Eq.\:\eqref{6}. For a large negative initial time $t_i\rightarrow-\infty$, if the system is initially in the instantaneous eigenstate with $\psi_1(-\infty)=0$ and $|\psi_2(-\infty)|=1$, we have $(S_1,S_2,S_0)=(0,0,1)$. Hence, we obtain a simple analytical formula for the total population at $t\rightarrow\infty$
\begin{align}\label{SimpleFormula1}
S_0(\infty)&\approx \cosh(2\Gamma \Delta t)+2\Gamma\sinh(2\Gamma\Delta t)\int_{t_2}^{\infty}\cos\Phi(t)dt\nonumber\\
&-4\Gamma\int_{t_1}^{t_2}v(t)\sinh(2\Gamma(t-t_1))dt\int_{t_2}^{\infty}\sin\Phi(t)dt,
\end{align}
where $\Delta t\equiv t_2-t_1$ is the size of the region of broken $\mathcal{P}\mathcal{T}$ symmetry. In particular, Eq.\:\eqref{SimpleFormula1} recovers the result $S_0(\infty)=1$ for $\Gamma=0$. In general, when there are more than two exceptional points, analytical approximations to the final transmission probabilities can also be obtained by neglecting either of the terms $2\Gamma S_0$ or $-2vS_2$ in Eq.\:\eqref{5b} in the regions of unbroken or broken $\mathcal{P}\mathcal{T}$ symmetry, and gluing the solutions at the boundaries of different regions.
\begin{figure}[tbp]
\begin{center}
\subfloat[$v(t)=t+t^3,\Gamma=1$\label{sfig:Fig_4c}]{%
  \includegraphics[width=0.49\columnwidth]{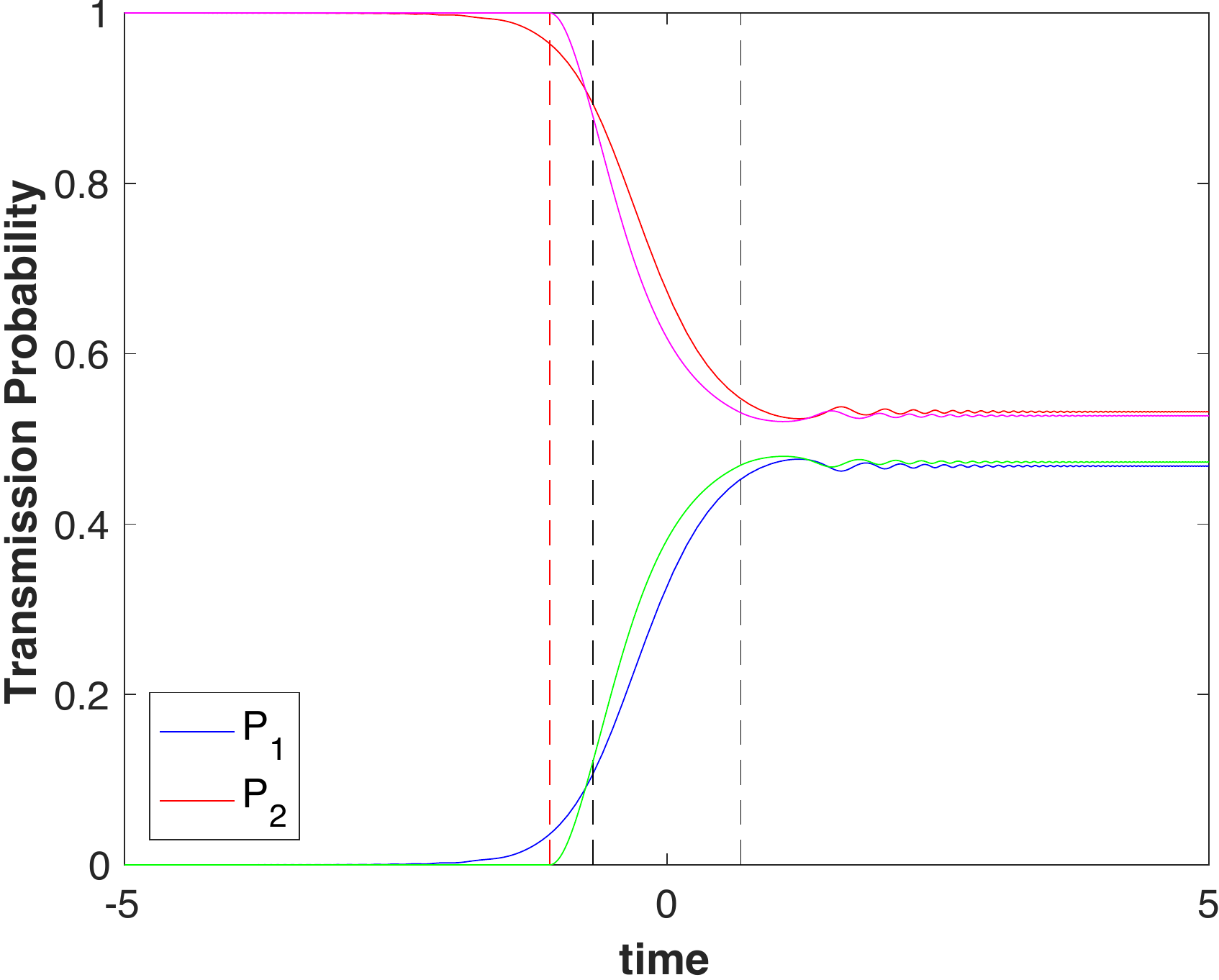}%
}\hfill
\subfloat[$v(t)=t+t^3,\Gamma=1$\label{sfig:Fig_4d}]{%
  \includegraphics[width=0.49\columnwidth]{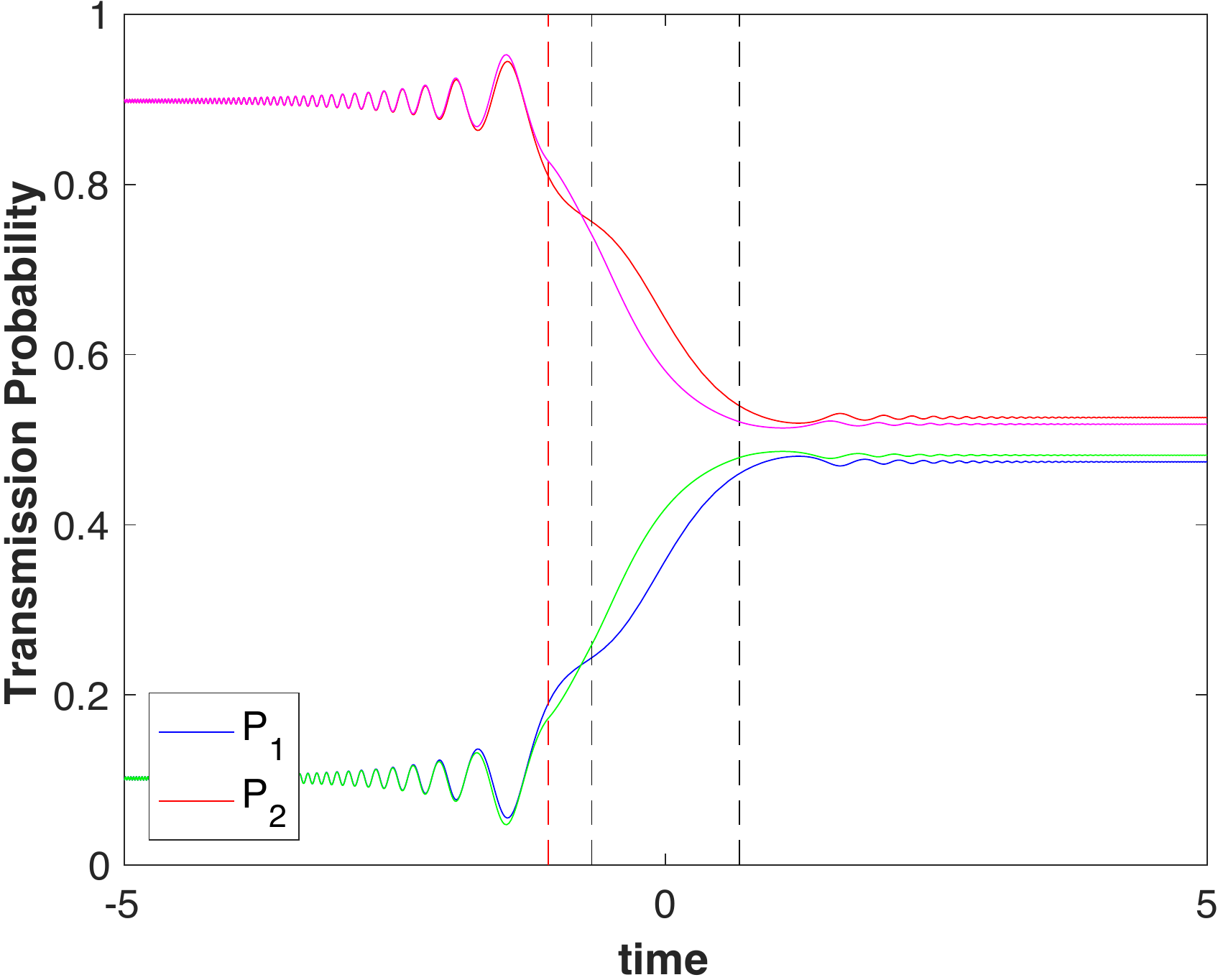}%
}
\caption{Schematic of the transmission probabilities for the super-parabolic case ($\alpha=\gamma=\Gamma=1$) with two exceptional points located at $t=\pm 0.68233$ (dashed lines). In Figs.\:\ref{sfig:Fig_4a} and \ref{sfig:Fig_4c}, the initial state is one of the instantaneous eigenstates; whereas in Figs.\:\ref{sfig:Fig_4b} and \ref{sfig:Fig_4d}, the initial state is a randomly selected state ($P_1=0.1$). The numerical solutions are depicted in red and blue lines; whereas the analytical approximate  solutions are depicted in magenta and green lines. In Figs.\:\ref{sfig:Fig_4c} and \ref{sfig:Fig_4d}, the first transition point $t_0=-1.08171$ is depicted in red dashed line.}
\label{fig:S0}
\end{center}
\end{figure}
In Fig.\:\ref{fig:tranS0beta}, we depict the transmission probabilities $P_1$ and $P_2$ based on the analytical formulas Eqs.\:\eqref{6} - \eqref{SimpleFormula1}, and compare the results to numerical simulations. Figs.\:\ref{sfig:Fig_6a} and \ref{sfig:Fig_6b} show that the final transmission probabilities are well-approximated by Eq.\:\eqref{SimpleFormula1} for the parabolic case for both initial instantaneous eigenstates and random selected initial states. However, as one may see from Figs.\:\ref{sfig:Fig_4a} and \ref{sfig:Fig_4b}, there is an overestimation of $P_2$ and an underestimation of $P_1$ for the super-parabolic case, which are possibly caused by the negligence of the contribution from the coupling between the two states just before reaching the first exceptional point. 

In order to reduce the accumulated errors in the final transmission probabilities, one may add a transition region in front of the first exceptional point. As one may see from Eqs.\:\eqref{dd1} - \eqref{dd2}, the term $v^2-\Gamma^2$ is exactly zero at the exceptional point and gradually increases until it balances the terms $\pm i\dot{v}$. Hence, the boundaries of the transition regions, which are referred to the transition points, may be determined by the condition $|\dot{v}(t)|=|v^2(t)-\Gamma^2|$. In particular, for $v(t)=\alpha t +\gamma t^3$, the transition points are the real roots of the sextic equation $\gamma^2t^6+2\alpha\gamma t^4+(\alpha^2\pm 3\gamma)t^2-\Gamma^2\pm\alpha=0$. Let us denote the transition point before the first exceptional point $t_1$ as $t_0$. For the transition region between $t_0$ and $t_1$, we may assume that $v\approx -\Gamma$, so that Eqs.\:\eqref{5a} - \eqref{5c} are replaced by $\dot{S}_0=2\Gamma S_1$, $\dot{S}_1=2\Gamma(S_0+S_2)$ and $\dot{S}_2=-2\Gamma S_1$. As a result, $S_0+S_2$ becomes a constant, and so does $\dot{S}_1$. Hence, the total population and the other two variables $S_1$ and $S_2$ can be approximated by
\begin{align}
    S_1(t)&\approx 2\Gamma(S_0(t_0)+S_2(t_0))t+S_1(t_0),\nonumber\\
    S_2(t)&\approx -2\Gamma^2(S_0(t_0)+S_2(t_0))t^2-2\Gamma S_1(t_0)t+S_2(t_0),\nonumber\\
    S_0(t)&\approx 2\Gamma^2(S_0(t_0)+S_2(t_0))t^2+2\Gamma S_1(t_0)t+S_0(t_0)\label{trans}.
\end{align}
Here, $S_k(t_0)=S_k(t_i)$ when the system is initially in an instantaneous eigenstate. For other cases, $S_k(t_0)$ are determined by Eq.\:\eqref{7} with $S_k(t_2)$ replaced by $S_k(t_i)$. In particular, for the special case that $\psi_1(-\infty)=0$ and $|\psi_2(-\infty)|=1$, we have $S_0(-\infty)=1$ and $S_1(-\infty)=S_2(-\infty)=0$. Hence, the modified analytical formula for the total population at $t\rightarrow\infty$ is
\begin{align}
    &S_0(\infty) \approx (1+2\Gamma^2t_1^2)\cosh(2\Gamma \Delta t)-2\Gamma t_1\sinh(2\Gamma\Delta t)\nonumber\\
    &+2\Gamma[(1+2\Gamma^2t_1^2)\sinh(2\Gamma \Delta t)-2\Gamma t_1\cosh(2\Gamma \Delta t)]\int_{t_2}^\infty\cos\Phi(t)dt\nonumber\\
    &-4\Gamma[(1+2\Gamma^2t_1^2)\int_{t_1}^{t_2}v(t)\sinh(2\Gamma(t-t_1))dt-\Gamma^2t_1^2\nonumber\\
    &+2\Gamma t_1\int_{t_1}^{t_2}v(t)\cosh(2\Gamma(t-t_1))dt]\int_{t_2}^\infty\sin\Phi(t)dt.
\end{align}
In Figs.\:\ref{sfig:Fig_4c} - \ref{sfig:Fig_4d}, the analytical approximations to the transmission probabilities are depicted for the super-parabolic case, after Eq.\:\eqref{trans} for $t_0\leq t\leq t_1$ are taken into account. The result shows that the final transmission probabilities are well-approximated by the modified analytical formulas for both initial instantaneous eigenstates and random selected initial states.

\section{Application to $\mathcal{PT}$-symmetric tight-binding lattice}\label{IV}
We now discuss how the parabolic and super-parabolic models studied in the last sections can be realized in a $\mathcal{PT}$-symmetric non-Hermitian one-dimensional tight-binding optical waveguide lattice with an index gradient, where the hopping dynamics of a single particle on the lattice is described by the Hamiltonian \cite{longhi2009bloch, garanovich2012light, della2013spectral, longstaff2019nonadiabatic, longhi2015robust, xu2016experimental, turker2016super}
\begin{equation}\label{OneWay}
\hat{H}\equiv \sum_{n}\{-\kappa(|n\rangle\langle n+1|+|n+1\rangle\langle n|)+[i\Gamma(-1)^n+Fn]|n\rangle\langle n|\},
\end{equation}
where $\kappa$ is the hopping rate between the adjacent sites, $\Gamma(-1)^n$ is an alternating gain and loss of the site energies, which may be achieved by metal-cladding on waveguides with $n$ odd, and $F$ is an index gradient along the lattice, which may be experimentally achieved by bending the waveguides \cite{xu2016experimental}. The one-dimensional lattice described by the Hamiltonian \eqref{OneWay} can be used to achieve one-way robust light transport in the present of disorder \cite{longhi2015robust, xu2016experimental}.

We now study the non-Hermitian system in the basis of the Bloch states $|k\rangle\equiv \frac{1}{\sqrt{2\pi}}\sum_ne^{ikn}|n\rangle$, where $k\in [-\pi,\pi]$ is the crystal momentum. In the absence of the index gradient ($F=0$), the Bloch state $\psi(k)$ in the crystal momentum representation obeys $i(d/dt)\psi(k)=-2\kappa\cos k\psi(k)+i\Gamma\psi(k+\pi)$, and the state $\psi(k+\pi)$ obeys $i(d/dt)\psi(k+\pi)=2\kappa\cos k\psi(k+\pi)+i\Gamma\psi(k)$. Hence, one may introduce the two-component state vector $|\Psi(k)\rangle\equiv(\psi(k),\psi(k+\pi))^T$, whose time evolution is governed by the Bloch Hamiltonian
\begin{equation}\label{NonHermiteBlochHamiltonian}
    \hat{h}(k) = \left( {\begin{array}{cc}
   -2\kappa\cos k & i\Gamma \\
   i\Gamma & 2\kappa\cos k \\
  \end{array} } \right).
\end{equation}

When a static force is applied to the lattice by engineering the refractive index of the waveguides, an initial state that is close to an eigenstate of the Bloch Hamiltonian \eqref{NonHermiteBlochHamiltonian} would experiences non-adiabatic transitions between the energy bands, which are non-Hermitian generalizations of the conventional Bloch oscillations, and correspond to a splitting of the beam in real space. In such a case, the Hamiltonian which governs the two-component state vector becomes
\begin{equation}
\hat{h}(k,q) = \left( {\begin{array}{cc}
   -2\kappa\cos k-Fq & i\Gamma \\
   i\Gamma & 2\kappa\cos k-Fq \\
  \end{array} } \right),
\end{equation}
where $q\equiv id/dk$ is canonical conjugate to $k$, i.e., $[q,k]=i$. In the Hermitian band theory, the expectation value of the crystal momentum obeys the acceleration theorem, $\langle k\rangle_t = \langle k\rangle_0 +Ft$. As shown by Longstaff and Graefe \cite{longstaff2019nonadiabatic}, the acceleration theorem can be applied to non-Hermitian systems, as long as the initial uncertainty in the crystal momentum is negligible. Hence, the non-adiabatic transition dynamics can be effectively described by the Hamiltonian \eqref{NonHermiteBlochHamiltonian}, with the crystal momentum $k$ being replaced by its expectation value $\langle k\rangle_t=\langle k\rangle_0+Ft$. One may then Taylor expand the effective Hamiltonian around the band edge $k=\pi/2$, and obtains (after shifting the time origin)
\begin{equation}\label{NewSuperParabolicHamiltonian}
    \hat{h}(t^\prime) \approx \left( {\begin{array}{cc}
   -\alpha t^\prime-\gamma t^{\prime 3} & i\Gamma \\
   i\Gamma & \alpha t^\prime+\gamma t^{\prime 3} \\
  \end{array} } \right),
\end{equation}
where $\alpha\equiv 2\kappa F$, $\gamma\equiv -\kappa F^3/3$, $t^\prime\equiv t-t_0$, and $t_0\equiv \frac{1}{2F}(\pi-2\langle k\rangle_0)$. The resulting Hamiltonian \eqref{NewSuperParabolicHamiltonian} is then equivalent to the non-Hermitian Hamiltonian $\eqref{NonHermiteHamiltonian}$ for the super-parabolic case. Interestingly, the number of exceptional points of the Hamiltonian \eqref{NewSuperParabolicHamiltonian} is irrelevant to the amplitude of the static force $F$. For the case that $\kappa F<0$ and $|\kappa|<3\Gamma/4\sqrt{2}$, there are two exceptional points; when $|\kappa|>3\Gamma/4\sqrt{2}$, there are six exceptional points; when $|\kappa|=3\Gamma/4\sqrt{2}$, there are four exceptional points. Similarly, one may Taylor expand the effective Hamiltonian around $k=0$, and obtains (after subtracting a constant $-2\kappa+\kappa\langle k\rangle_0^2$ from the Hamiltonian)
\begin{equation}\label{NewParabolicHamiltonian}
    \hat{h}(t) \approx \left( {\begin{array}{cc}
   -\alpha t-\beta t^2 & i\Gamma \\
   i\Gamma & \alpha t+\beta t^2 \\
  \end{array} } \right),
\end{equation}
where $\alpha\equiv -2\kappa\langle k\rangle_0 F$ and $\beta\equiv -2\kappa F^2$. The resulting Hamiltonian \eqref{NewParabolicHamiltonian} is then equivalent to the non-Hermitian Hamiltonian $\eqref{NonHermiteHamiltonian}$ for the parabolic case. Similar to the super-parabolic case, the number of exceptional points for Hamiltonian \eqref{NewParabolicHamiltonian} is irrelevant to the amplitude of the static force $F$. For $|\kappa|<2\Gamma/\langle k\rangle_0^2$, there are two exceptional points; for $|\kappa|>2\Gamma/\langle k\rangle_0^2$, there are four exceptional points; for $|\kappa|=2\Gamma/\langle k\rangle_0^2$, there are three exceptional points.

\section{Conclusion}\label{V}
We discussed the non-Hermitian dynamics of a two-level quantum system driven through an assembly of exceptional points at finite speed which are quadratic or cubic functions of time. We derived analytical approximate formulas for the non-adiabatic transmission probabilities for both the parabolic and super-parabolic cases. We demonstrated possible experimental realizations in one-dimensional graded index photonic crystal waveguide, which may be applied to unidirectional light transport in modulated waveguides. We found that for both the parabolic and super-parabolic cases, the number of exceptional points increases as the hopping rate between the neighboring sites increases. In future works, we may extent the current approximation procedure to the cases where both the amplitude of the alternating gain and loss and the index gradient are time dependent.   

\begin{acknowledgements}
The Authors would like to thank the Science and Technology Development Fund of the Macau SAR for providing support, FDCT 023/2017/A1.
\end{acknowledgements}

\begin{appendix}
\section{Evaluation of Eq.\:\eqref{SimpleFormula1} for the parabolic model}
For $v(t)\equiv\alpha t+\beta t^2$ with $\alpha^4<16\Gamma^2\beta^2$, the integral which involves hyperbolic sine function in Eq.\:\eqref{SimpleFormula1} may be evaluated using the indefinite integral
\begin{align}
&\int v(t)\sinh(2\Gamma(t-t_1))dt \nonumber\\
&= (\frac{v}{2\Gamma}+\frac{\ddot{v}}{8\Gamma^3})\cosh (2\Gamma(t-t_1))-\frac{\dot{v}}{4\Gamma^2}\sinh(2\Gamma(t-t_1)),
\end{align}
which yields
\begin{align}\label{I1}
I_1&\equiv\int_{t_1}^{t_2} v(t)\sinh(2\Gamma(t-t_1))dt \nonumber\\
&= (\frac{v(t_2)}{2\Gamma}+\frac{\beta}{4\Gamma^3})(\cosh(2\Gamma\Delta t)-1)-\frac{\dot{v}(t_2)}{4\Gamma^2}\sinh(2\Gamma\Delta t)\nonumber\\
&=  (\pm\frac{1}{2}+\frac{\beta}{4\Gamma^3})\left[\cosh\left(\frac{2\Gamma}{\beta}\sqrt{\alpha^2\pm4\beta\Gamma}\right)-1\right]\nonumber\\
&-\frac{\sqrt{\alpha^2\pm4\beta\Gamma}}{4\Gamma^2}\sinh\left(\frac{2\Gamma}{\beta}\sqrt{\alpha^2\pm4\beta\Gamma}\right),
\end{align}
where we have used $v(t_2)=v(t_1)=\pm\Gamma$, $\dot{v}(t_2)=\pm\sqrt{\alpha^2\pm4\beta\Gamma}$ and $\Delta t=\pm\frac{1}{\beta}\sqrt{\alpha^2\pm4\beta\Gamma}$ for $\beta=\pm|\beta|$. In particular, for $\Gamma\rightarrow 0$, we have $I_1\rightarrow \pm \Gamma^2\alpha^2/\beta^2$. 

We now evaluate the two definite integrals $\int_{t_2}^\infty\cos\Phi(t)dt$ and $\int_{t_2}^\infty\sin\Phi(t)dt$ in Eq.\:\eqref{SimpleFormula1}, where $\Phi(t)\equiv 2\int_{t_2}^tv(s)ds$. To begin with, let us consider the integral
\begin{equation}\label{Integral1}
I_2\equiv \int_{t_2}^\infty e^{i\Phi(t)}dt = e^{-i(\alpha t_2^2+\frac{2}{3}\beta t_2^3)} \int_{t_2}^\infty e^{i(\alpha t^2+\frac{2}{3}\beta t^3)}dt.
\end{equation}
After the change of variable $\tau\equiv t+\frac{\alpha}{2\beta}$, Eq.\:\eqref{Integral1} becomes
\begin{equation}\label{Integral2}
I_2= e^{-\frac{2i\beta}{3}(\tau_2^3-3k^2\tau_2)} \int_{\tau_2}^\infty e^{\frac{2i\beta}{3}(\tau^3-3k^2\tau)}d\tau,
\end{equation}
where $k\equiv\frac{\alpha}{2\beta}$. After the transformation $x\equiv (\frac{2\beta}{3})^{1/3}\tau$ and $\lambda\equiv 3k^2(\frac{2\beta}{3})^{2/3}$, Eq.\:\eqref{Integral2} becomes
\begin{align}\label{Integral3}
I_2&= (\frac{3}{2\beta})^{1/3}e^{-i(x_2^3-\lambda x_2)}\int_{x_2}^\infty e^{i(x^3-\lambda x)}dx\nonumber\\
&=(\frac{3}{2\beta})^{1/3}e^{-i(x_2^3-\lambda x_2)}\sum_{m=0}^\infty\frac{(-i\lambda)^m}{m!}\int_{x_2}^\infty x^m e^{ix^3}dx.
\end{align}
We now evaluate the generalized Fresnel integrals in terms of the confluent hypergeometric functions
\begin{align}
\int x^m e^{ix^n}dx &= \frac{x^{m+1}}{m+1}\setlength\arraycolsep{1pt}
{}_1 F_1\left(\begin{matrix}\frac{m+1}{n}\\\frac{m+n+1}{n}\end{matrix}\left.\right|ix^n\right)\nonumber\\
&=\frac{x^{m+1}}{m+1}e^{ix^n}\setlength\arraycolsep{1pt}
{}_1 F_1\left(\begin{matrix}1\\\frac{m+n+1}{n}\end{matrix}\left.\right|-ix^n\right).
\end{align}
Using the asymptotic expansion of the confluent hypergeometric function
\begin{equation}
\frac{x^{m+1}}{m+1}\setlength\arraycolsep{1pt}
{}_1 F_1\left(\begin{matrix}\frac{m+1}{n}\\\frac{m+n+1}{n}\end{matrix}\left.\right|ix^n\right)
\sim\frac{1}{n}\Gamma\left({\frac{m+1}{n}}\right)e^{i\pi(m+1)/(2n)}.
\end{equation}
Eq.\:\eqref{Integral3} becomes
\begin{align}\label{I2}
I_2&=(\frac{3}{2\beta})^{1/3}\frac{e^{i\pi/6}}{3}e^{-i(x_2^3-\lambda x_2)}\sum_{m=0}^\infty\frac{(-i\lambda e^{i\pi/6})^m}{m!}\Gamma\left({\frac{m+1}{3}}\right)\nonumber\\
&-(\frac{3}{2\beta})^{1/3}e^{i\lambda x_2}\sum_{m=0}^\infty\frac{(-i\lambda x_2)^m}{m!}\frac{x_2}{m+1}\setlength\arraycolsep{1pt}
{}_1 F_1\left(\begin{matrix}1\\\frac{m+4}{3}\end{matrix}\left.\right|-ix_2^3\right).
\end{align}
Substituting Eqs.\:\eqref{I1} and \eqref{I2} into Eq.\:\eqref{SimpleFormula1}, the total population at $t\rightarrow \infty$ becomes
\begin{align}
S_0(\infty)&\approx\cosh\left(\frac{2\Gamma}{\beta}\sqrt{\alpha^2\pm4\beta\Gamma}\right)\pm2\Gamma\sinh\left(\frac{2\Gamma}{\beta}\sqrt{\alpha^2\pm4\beta\Gamma}\right)\cdot\nonumber\\
&(\frac{3}{2\beta})^{1/3}\mbox{Re}\left(\frac{e^{i\pi/6}}{3}e^{-i(x_2^3-\lambda x_2)}\sum_{m=0}^\infty\frac{(-i\lambda e^{i\pi/6})^m}{m!}\Gamma\left({\frac{m+1}{3}}\right)\right.\nonumber\\
&\left.-e^{i\lambda x_2}\sum_{m=0}^\infty\frac{(-i\lambda x_2)^mx_2}{m!(m+1)}\setlength\arraycolsep{1pt}
{}_1 F_1\left(\begin{matrix}1\\\frac{m+4}{3}\end{matrix}\left.\right|-ix_2^3\right)\right)\nonumber\\
&-\left[(\pm4\Gamma+\frac{\beta}{\Gamma^2})\left(\cosh\left(\frac{2\Gamma}{\beta}\sqrt{\alpha^2\pm4\beta\Gamma}\right)-1\right)\right.\nonumber\\
&\left.-\frac{\sqrt{\alpha^2\pm4\beta\Gamma}}{\Gamma}\sinh\left(\frac{2\Gamma}{\beta}\sqrt{\alpha^2\pm4\beta\Gamma}\right)\right]\cdot\nonumber\\
&(\frac{3}{2\beta})^{1/3}\mbox{Im}\left(\frac{e^{i\pi/6}}{3}e^{-i(x_2^3-\lambda x_2)}\sum_{m=0}^\infty\frac{(-i\lambda e^{i\pi/6})^m}{m!}\Gamma\left({\frac{m+1}{3}}\right)\right.\nonumber\\
&\left.-e^{i\lambda x_2}\sum_{m=0}^\infty\frac{(-i\lambda x_2)^mx_2}{m!(m+1)}\setlength\arraycolsep{1pt}
{}_1 F_1\left(\begin{matrix}1\\\frac{m+4}{3}\end{matrix}\left.\right|-ix_2^3\right)\right).
\end{align}

\end{appendix}

\end{document}